\documentclass[acmsmall]{acmart}
\AtBeginDocument{%
  }

\usepackage{graphicx}
\usepackage{subcaption}
\usepackage{wrapfig}
\usepackage{multirow}
\graphicspath{{figures/}}
\usepackage[dvipsnames]{xcolor}
\usepackage[most]{tcolorbox}

\tcbset{
  badge/.style={
    on line, 
    boxsep=3pt,
    left=0pt,
    right=0pt,
    top=0pt,
    bottom=0pt,
    colframe=white,
    fontupper=\color{white}
  }
}

\newtcbox{\mirrorBadge}{
  badge,
  colback=RubineRed
}

\newtcbox{\transpBadge}{
  badge,
  colback=BlueViolet
}

\newtcbox{\scaffoldBadge}{
  badge,
  colback=Mahogany
}

\newtcbox{\flexiBadge}{
  badge,
  colback=BurntOrange
}

\newtcbox{\relateBadge}{
  badge,
  colback=Fuchsia
}

\begin{document}

\title{Voice to Vision: Enhancing Civic Decision-Making through Co-Designed Data Infrastructure}


\author{Margaret Hughes}
\authornote{All authors contributed equally to this research.}

\affiliation{%
  \institution{Massachusetts Institute of Technology}
  \city{Cambridge}
  \state{MA}
  \country{USA}}
  \email{mhughes4@media.mit.edu}

\author{Cassandra Overney}
\affiliation{%
  \institution{Massachusetts Institute of Technology}
  \city{Cambridge}
  \state{MA}
  \country{USA}}
\authornotemark[1]
\email{coverney@mit.edu}

\author{Ashima Kamra}
\affiliation{%
  \institution{Wellesley Collegey}
  \city{Wellesley}
  \state{MA}
  \country{USA}}

\author{Jasmin Tepale}
\affiliation{%
  \institution{New York City Department of City Planning}
  \city{New York City}
  \state{NY}
  \country{USA}}

\author{Elizabeth Hamby}
\affiliation{%
  \institution{New York City Department of City Planning}
  \city{New York City}
  \state{NY}
  \country{USA}}

\author{Mahmood Jasim}
\affiliation{%
  \institution{Louisiana State University}
  \city{Baton Rouge}
  \state{LA}
  \country{USA}}

\author{Deb Roy}
\affiliation{%
  \institution{Massachusetts Institute of Technology}
  \city{Cambridge}
  \state{MA}
  \country{USA}}

\renewcommand{\shortauthors}{Hughes \& Overney \& Kamra \& Tepale \& Hamby \& Jasim \& Roy}

\begin{abstract}
Trust and transparency in civic decision-making processes, like neighborhood planning, are eroding as community members frequently report sending feedback ``into a void'' without understanding how, or whether, their input influences outcomes. 
To address this gap, we introduce Voice to Vision, a sociotechnical system that bridges community voices and planning outputs through a structured yet flexible data infrastructure and complementary interfaces for both community members and planners. 
Through a five-month iterative design process with 21 stakeholders and subsequent field evaluation involving 24 participants, we examine how this system facilitates shared understanding across the civic ecosystem. 
Our findings reveal that while planners value systematic sensemaking tools that find connections across diverse inputs, community members prioritize seeing themselves reflected in the process, discovering patterns within feedback, and observing the rigor behind decisions, while emphasizing the importance of actionable outcomes. 
We contribute insights into participatory design for civic contexts, a complete sociotechnical system with an interoperable data structure for civic decision-making, and empirical findings that inform how digital platforms can promote shared understanding among elected or appointed officials, planners, and community members by enhancing transparency and legitimacy.
\end{abstract}



\received{20 February 2007}
\received[revised]{12 March 2009}
\received[accepted]{5 June 2009}
\begin{teaserfigure}
  \centering
  \includegraphics[width=\textwidth]{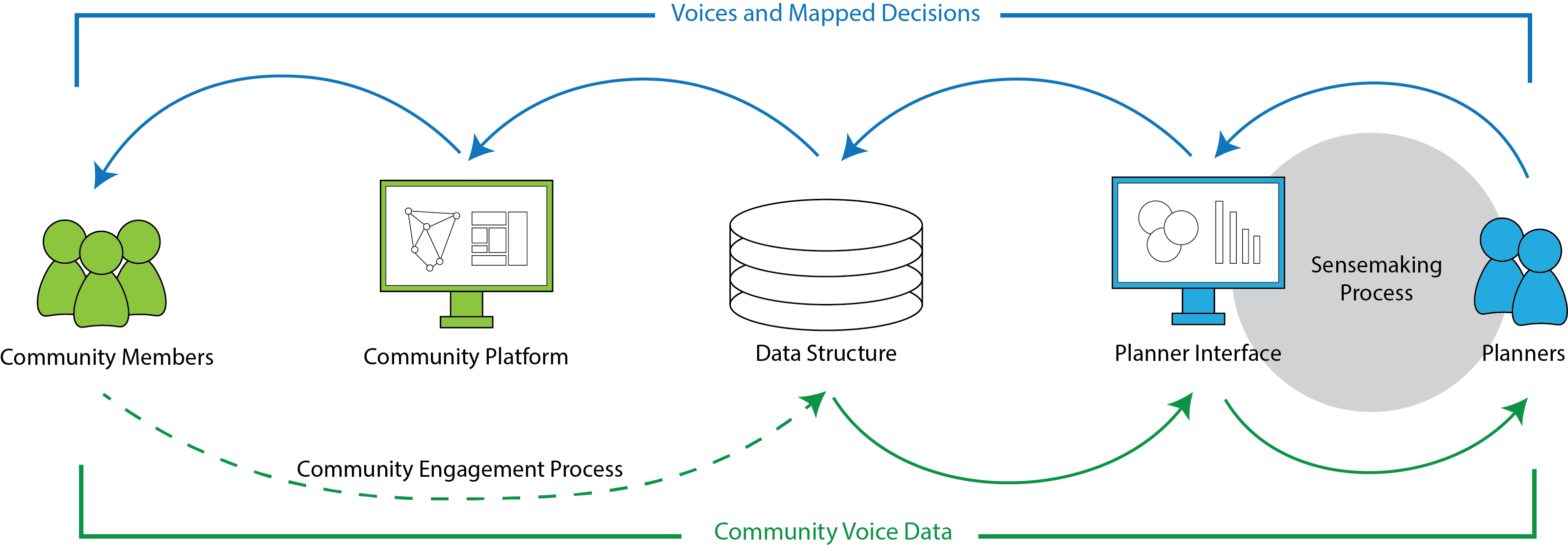} 
  \caption{A diagram of the Voice to Vision system showing the flow of community voice data between community members and planners during an engagement process. This system helps close feedback loops between both stakeholder groups.}
  \Description{TODO}
\end{teaserfigure}
\maketitle

\vspace{-0.2cm}
\section{Introduction}

Trust and legitimacy in civic processes are foundational to effective democratic governance, yet in urban planning and other civic domains, a persistent challenge exists in connecting community input to decision outcomes. 
While planners invest significant resources in gathering community voices, the process of transforming this input into concrete planning outputs remains opaque, often leading to perceptions that engagement efforts merely ``check a box'' rather than meaningfully shape decisions~\cite{overney2025boundarease}.
The challenge of building shared understanding between planners and communities is particularly acute in metropolitan areas, where neighborhoods may house populations equivalent to small cities (15,000-100,000 residents). 
At this scale, cooperation requires navigating diverse current needs against future priorities while considering a multitude of hopes, concerns, and visions~\cite{baumerCourseItPolitical2022}. 
These complex planning processes involve multiple stakeholder groups, such as planners, community members, other city agencies, and elected or appointed officials, each with distinct needs, constraints, and perspectives on what constitutes successful engagement~\cite{jasimCommunityPulseFacilitatingCommunity2021}. 
Planning departments increasingly face expectations to demonstrate responsiveness to community input, creating a significant need for systems that support both rigorous sensemaking of feedback and transparent communication of outcomes~\cite{villanueva2017bringing}.

Addressing these complex cooperation challenges requires sociotechnical solutions that balance technological capacity with relational sensitivity. 
In HCI and CSCW fields, researchers have developed technological systems to improve both the gathering of community input~\cite{overney2025coalesce, jasimCommunityClickCapturingReporting2021, kripleanSupportingReflectivePublic2012} and the analysis capabilities of decision-makers~\cite{jasimCommunityPulseFacilitatingCommunity2021}. 
In addition, platforms like Polis~\cite{smallPolisScalingDeliberation2021} and other deliberative systems demonstrate the potential for technology to facilitate large-scale civic participation.
Within planning practice, considerable work focuses on strengthening the social relationship-building aspects of community engagement~\cite{harvey1970social}. 

However, a critical gap persists at the intersection of planning practice and CSCW research.
While each domain offers valuable insights, they rarely inform each other.
Planning literature emphasizes relationship-building without technological sophistication, while civic technology often adopts transactional approaches that focus narrowly on either collecting input or supporting analysis in isolation~\cite{corbettProblemCommunityEngagement2018}. 
This separation has resulted in few systems that create comprehensive data infrastructures connecting community voices to planning outputs. 
Such fragmentation contributes to the black box perception of planning, where residents provide feedback but cannot trace how (or whether) it influences outcomes, further eroding trust in civic processes. 
Our work addresses this gap through an integrated sociotechnical system that supports both the analytical needs of planners and the transparency demands of community members.

We conducted a five-month iterative design process, employing co-design and participatory design methods with 12 planners (two of whom are co-authors) and 9 community members.
This process identified 5 key design goals that informed the development of Voice to Vision, a comprehensive sociotechnical system comprising a data structure and two complementary interfaces: a sensemaking interface for planners and a community-facing platform for the public. 
We evaluated the system through a six-week field deployment of the community-facing platform in an ongoing neighborhood planning process with 15 community member interviews.
The sensemaking interface was evaluated through qualitative user studies with 7 planners. 

Our evaluation revealed Voice to Vision's potential to strengthen the connection between community input and planning outcomes. 
Planners emphasized the system's value in supporting systematic sensemaking by enabling them to find connections across diverse inputs and communicate outputs accessibly, a capability lacking in their typical spreadsheet-based workflows. 
Among community members, we observed multifaceted benefits: participants valued seeing themselves and others reflected in the voices, discovering patterns within community feedback, and witnessing the rigor behind decision-making processes. 
However, they consistently emphasized actionability as a critical concern, wanting to see concrete paths from community input to implementation. 
Both stakeholder groups demonstrated a preference for digestible information formats while appreciating the availability of deeper exploration opportunities. 
Perhaps most significantly, the platform facilitated a sense of legitimacy and care across the civic ecosystem, helping build shared understanding between officials, planners, and community members.
The primary contributions of our work are:
\begin{enumerate}
    \item Insights and reflections on a design process with novel mitigations to common challenges that arise when doing participatory design in governance contexts
    \item Voice to Vision, a complete sociotechnical system with an interoperable data structure for urban planning, community engagement, and constituency-informed decision-making
    \item Empirical insights from a field deployment and qualitative interviews with 24 participants that inform future work on designing tools for civic decision-making to promote shared understanding between stakeholders
\end{enumerate}

\subsection{Positionality Statement}

Our research team brings together academic researchers from human-computer interaction and data visualization with practicing urban planners who have direct responsibility for neighborhood planning processes. 
Two of our co-authors are city planners who simultaneously played multiple roles during this work: they managed the specific planning process studied, contributed to department-wide process improvements, and represented a public agency with civic leadership responsibilities. 
Additionally, one co-author is a resident of the neighborhood where we conducted our field deployment, bringing both researcher and community member perspectives to the work.
We acknowledge that our team's shared interest in creating more transparent connections between community voices and planning outputs influenced our design priorities, sometimes prioritizing this challenge over other concerns raised by community members during our design process. 
We used co-design methods with planners, integrating them fully into research and design decisions, while employing participatory design with community members through workshops and design sessions. 
This differential engagement reflects our strategic emphasis on partnering with planners who could implement systemic changes across planning processes. 
While we strived to incorporate diverse community perspectives, we recognize that our approach may have privileged institutional knowledge over community expertise at times. 
Our positionality as researchers working closely with government practitioners shaped both the strengths and limitations of the resulting sociotechnical system.

\section{Background}


\subsection{Participatory Planning as a Democratic Practice} 

\subsection{Design Methodologies in Civic Technology and Digital Civics} 

\subsection{HCI Systems for Civic Decision Making}

\section{Design Methodology}

This section describes a five-month iterative design process involving co-design and participatory design methods with 12 planners and 9 community members.






\subsection{High-Level Design Framework}

To develop a system that facilitated shared understanding between a constituency and their decision makers, it was essential that we engaged with both groups. As outlined, work in civic technology leverages participatory design methods techniques to generate knowledge and create designs responsive to the unique conditions and needs of the users \cite{vlachokyriakos2016digital, le2015strangers, le2009values}. In values-sensitive design, there is extensive documentation about the values-negotiations that occur within these design processes, as the partners we work with are not a monolith, and various values will emerge in tension with one another. 

\begin{figure}[t]
    \centering
    \includegraphics[width=0.75\linewidth]{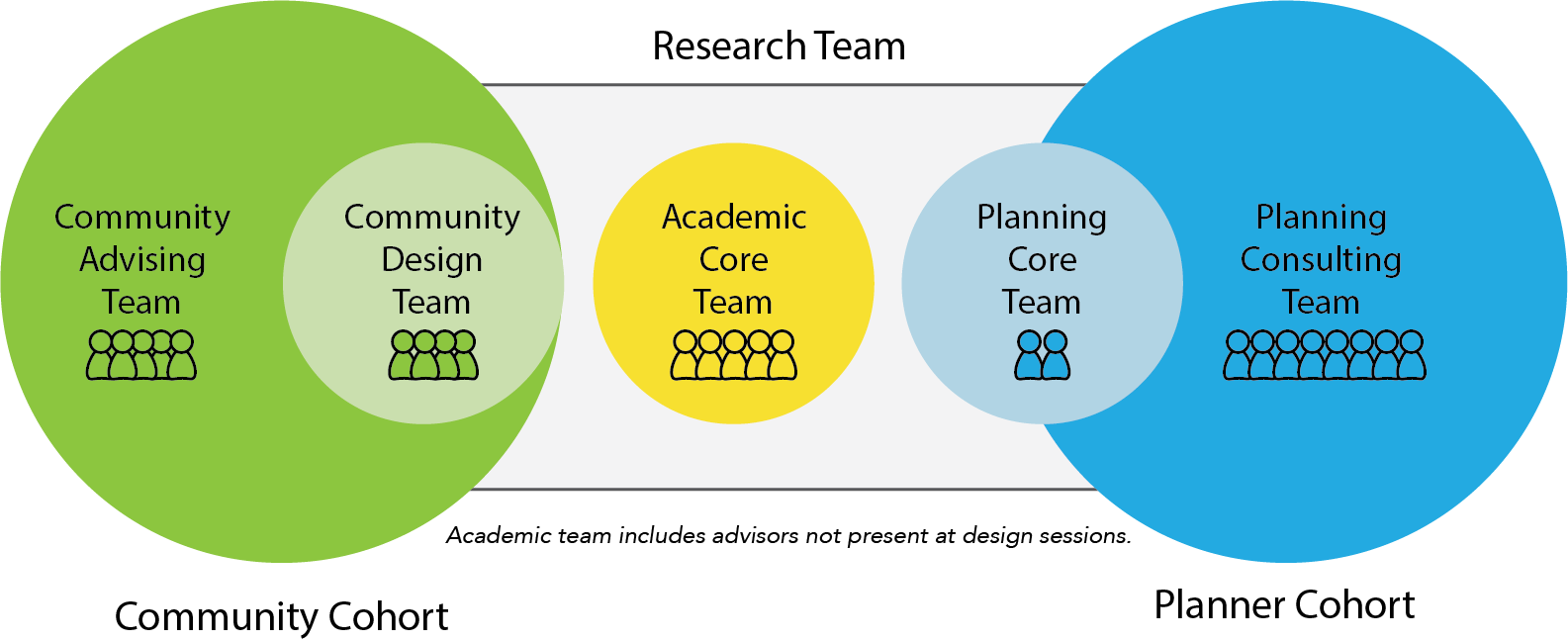}
    \caption{A diagram of our design methodology involving planners and community members. Each design cohort has concentric circles representing different levels of involvement.\vspace{-0.5cm}}
    \label{fig:design-process}
    \Description{TODO}
\end{figure}

We approached this design challenge by designing with two distinct groups, employing different engagement methods with each to develop and refine our designs (Figure~\ref{fig:design-process}). We recruited a community cohort of 9 participants, and a planner cohort interested in community engagement methods and tools. Our research team, including the core planner team, bridged the two groups, designing workshops for each team. 

\subsubsection{Partnering with Civic Leaders}

As we were particularly interested in voice-to-decision flow, we decided to identify civic leaders who had a forthcoming constituency-informed decision requirement. We sent out a broad call to civic leaders, and from that call, a planning team from a large coastal city in the United States responded.  We decided to work on a specific neighborhood plan with that planning team, and then bring together a community cohort from the impacted neighborhood to design the community facing platform. This plan was already in development and a majority of community engagement had already been completed, so our focus was primarily on how to bring that data into the Voice to Vision system with input from planners and community members. The academic team, the director of civic engagement, and the neighborhood plan proejct manager formed the research team (Figure~\ref{fig:design-process}), and all are authors on this paper. Both the community and planners are essential partners in this process, yet the responsibility of decision making and communicating falls upon the planning team; therefore the academic team partnered most closely with planners. The resulting research team met for at least one hour each week from summer to spring of 2024 and co-designed the research processes.

\subsubsection{Partnering with Community Members}
As our project explores how to make civic decision-making processes more transparent and trustworthy, a key impacted group is the community for whom the outcome of these decisions would most impact. As we were designing for a particular planning process, it was necessary to get community input from the planning study area. We decided to recruit a cohort and engage in participatory design. 
Participatory methods enable a deeper understanding of the problems at hand, especially from the lenses and experiences of those most often marginalized in civic processes. By partnering closely with these groups, we would be able to more deeply understand what values felt most important and urgent to the community \cite{muller1993participatory}.

\subsubsection{Design Cohort Structure}
Both design cohorts were diverse with about 10 people in each. With this size, not all members could commit to the level of partnership needed. Further, with this size and diversity, different participants had different strengths, interests, and availabilities. So, we developed a novel nested sub-team structure to facilitate the design process, as shown in Figure~\ref{fig:design-process}. With this approach, we created sub-teams within each cohort that consisted of 2 members of the planning cohort who are co-authors of this paper and co-leads throughout the whole project, and a sub-team of 4 community members who regularly attended design meetings to make design decisions together. The larger teams for each cohort acted as advisors, guiding the overall project direction, while the sub-teams acted as designers, contributing to decision-making, meeting more regularly, and helping bridge with the larger group. Therefore, we had a smaller \textit{design or core team} nested within a larger \textit{advising or consulting team} for each cohort.

\subsection{Planner Design Process}
Planners were selected to represent the diversity of teams across the lead agency, as well as partner agencies. They included staff from different functional roles, as well as geographic focus areas.
For the planning cohort, we had a group of 12 planners (Table~\ref{tab:plannercohort}) who participated in one-on-one interviews for initial information gathering, and then proceeded to complete 2 design sessions, often with 1-2 planners in each design session and 1-2 members of the academic team. 
All interview protocols can be found in the supplementary materials (S1-S3).

\subsubsection{Informational Interviews}
For the interviews, we took a semi-structured approach. We started the interview by inviting participants to describe a community engagement project they have worked on. We invited them to reflect on how they gathered voices, how they analyzed them, what they learned. We further asked about they tools they used, what challenges and smooth points existed in this process, and what questions they hoped to answer from the data. 
We analyzed notes from the interviews with affinity diagramming~\cite{lucero2015using, holtzblatt2005rapid}, a recommended approach for design evaluations that facilitates identification of patterns across participant feedback. 
This process involved segmenting transcripts into notes and iteratively clustering them to identify themes. 
From two clustering iterations, we developed an initial understanding of the scope of planning processes and the diversity and similarity of methods used across a particular city. We also identified a series of pain points and opportunities for intervention.

\subsubsection{User Interviews}
After the informational interviews, we held 2 rounds of user interviews. We focused on how to respond to key community engagement pain points, such as how to create new projects from scratch and how to explore insights. We performed rapid prototyping to create a series of low-fidelity interfaces through Figma\footnote{\url{https://www.figma.com/}.} to walk through specific scenarios and challenges with the planners. Throughout the session, we facilitated discussions and prompted around their specific interests in each interface idea. 
During these interviews, we also developed and iterated upon a data structure that would connect with any prototyped interfaces. We offered an initial structure based on the informational interviews and refined it with the planners. We identified what was missing, what fields should be more flexible or more rigid, and what controls planners have over the publicity of certain components. 
.


\subsection{Community Design Process}


We had an open call to the community where we asked questions about why they were interested in participating and about their experience in the planning process. We had about 50 applicants. After reaching out and offering 10 spots to applicants, 9 accepted and attended our first community workshop (Table~\ref{tab:communitycohort}). Most participants identified as African-American women, with one South Asian participant, although there was a strong South Asian presence in the community. This skew garnered critique from the participants themselves, for not having men's voices or more participants from various ethnicities.
We tried to respond to this critique in subsequent workshops through persona exercises and in our evaluation methods. Of those 9, three spots were held and reserved for key community leaders in the neighborhood who were very vocal at town halls and held a great deal of social capital. These spots were reserved at the advice of the neighborhood's city councilmember for community buy-in. These voices had a strong impact on the group dynamics, which we elaborate on in Section ~\ref{sec:power}.
All workshop protocols can be found in the supplementary materials (S4-S8).








\subsubsection{Long-Form Workshops}
We held two long-form workshops in person in the neighborhood involving the entire community cohort, which lasted from 10 am to 4:45 pm. The workshops were designed iteratively over several weeks, and one research team member was responsible for extensive note taking. We arranged for local foods and snacks as refreshments, and offered participants compensation with Visa gift cards for \$20 per hour. In these two long-form workshops, the academic and planner core teams co-facilitated. We chose to have this cross-over of cohorts at this stage because the planning side of our team holds a deep understanding of the plan, the planning process, and the neighborhood, and offered extensive expertise in community engagement and facilitation. We grounded the workshops in exercises to identify values and needs with the community and norms for how we should work together. We returned to these values and norms at moments of disagreement and when norms were broken. The workshops included dialogue and story-sharing, as well as design activities such as persona creation, working with qualitative data cards, and iterating on visualizations. At the end of the workshops, we held and took notes on what worked or could be improved, and asked for key takeaways. After each workshop, the research team held retrospective reflection meetings to synthesize learnings and design the next workshop.

\subsubsection{Power Dynamics within Workshops}\label{sec:power}
During the long-form workshops and when reviewing the workshop notes, we observed multifaceted power dynamics that influenced participation and discourse. In line with the literature, we observed that the community leaders with reserved spots, often the older participants with more extensive civic engagement experience, sometimes dominated discussions, creating an imbalance that affected the distribution of voice. 
Conversely, younger participants demonstrated more technological knowledge and comfort, creating a different power axis within design activities focused on digital tools. These intersecting dimensions of power required careful facilitation to ensure all voices were heard and valued, including returning to our norms and values repeatedly and leaning on community members to hold one another accountable to those norms. Further, some participants had particular strengths and voices that may not shine in more diverse groups, so we shifted towards small group collaborations, balancing between the richness and value expressed from the cohort when there was cross-generational collaboration and the greater equality of participation that occurred in more homogeneous groups. This learning also informed the split between the \textit{advising team} and the \textit{design team} enabling those who wish to do more technical design work the space to do so, and those who wish to advise or give insight into the broader scope of the project a venue to do so, while still being connected as a broader cohort.

\subsubsection{Virtual Workshops}
After the long-form workshops, due to financial limitations on the research team's side to travel to the site, we held two online Zoom workshops that continued the discussions of our in-person workshops, specifically on visualization design. 
Then, we split the teams into the \textit{design team} and \textit{advising team}, and proceeded to hold bi-weekly design sessions with the 4 member \textit{design team} for 5 sessions in total. Usually these sessions lasted 60 to 90 minutes, and we compensated participants \$20 for their time. Sessions consisted of iterative design work, such as workshoping designs on Figma and testing prototypes.
We invited the \textit{advising team} to give feedback on the platform during our final evaluation.


\section{Design Goals}
From this design process and related work 
emerged a set of five design goals that support the pursuit of shared understanding and cooperation between the community and urban planners. 
Design goals were determined through affinity diagramming~\cite{lucero2015using, holtzblatt2005rapid} of workshop notes, along with notes from retrospective reflections among the research team.

\subsection{\mirrorBadge{DG-Mirror} Connection and Mirroring }
Community voice serves as a valuable collective asset, yet engagement processes often extract feedback without returning findings to residents \cite{gor}. 
Our research revealed both planners and community members desire greater visibility of diverse perspectives within neighborhoods. 
As C10 observed when exploring voices that contradicted local conversations about housing priorities: \textit{``When people talk about housing in this community, it's always related to affordable housing and apartment buildings. A lot of people talked about homeowners and taxes... which flies in the face of what we've been hearing for the last 3 years.''} 
Community members valued seeing these differences reflected back, noting that \textit{``goals of the plan might not be the goals of the community,''} (C22) while simultaneously desiring unity without concealing disagreements. 
Meanwhile, planners expressed tension around who speaks for a neighborhood, with some valuing demographic context because it is \textit{``helpful to understand who is giving the feedback''} (P3) while others warned of the \textit{``ugly undercurrent of weeding out people's voices''} (P2) based on identity characteristics. 
These insights led to our design goal: \textit{Make community voices, and the various groups, interconnections, and differences within them, visible.}

\subsection{\transpBadge{DG-Transp} Transparency}
Civic engagement processes often suffer from perceptions that community feedback disappears into a void without meaningfully influencing decisions, a challenge well-documented in literature\cite{arnsteinLadderCitizenParticipation1969, gordon2014playful} and confirmed in our research. 
Our design process revealed deteriorating trust between community members and planners, with participants expressing skepticism about data collection methods: \textit{``A deep dive needs to be looked into how the data is collected and what questions are asked and why we are getting the data we are getting''} (C20). 
Planners simultaneously faced technical limitations when processing qualitative data while desiring to communicate their analytical rigor: \textit{``We currently don't have a systematic approach... What I would love is some systematic way to say `these are all the insights we have,' then `here are recommendations'''} (P16). 
Community members specifically wanted greater connection between their input and outcomes: \textit{``I want to see the comparison of what the community is saying and what is actually being done to see if community concerns are actually being accounted for''} (C5), highlighting the critical relationship between transparency and perceived legitimacy. 
These findings led to our design goal: \textit{Make community voices and their impact on decisions transparent.}

\subsection{\scaffoldBadge{DG-Scaffold} Accessibility and Scaffolding} 
Repeated throughout the community cohort sessions and previous literature is the importance of accessibility and scaffolding when developing interfaces that display community voice data. The planning cohort highlighted how difficult of a task this can be when communicating complex engagement processes: \textit{``there isn't one right way to do it... Principles for how we approach putting information as much as possible: as much plain language as possible, as straightforward and relevant to the topic as possible, don't bombard people with everything about the topic, [and] visuals help''} (P1). Beyond just communicating the information in an accessible way, we needed to ensure the system works across devices with multiple language options while applying basic principles of accessible and inclusive interface design (Mace et al., 1991). The importance of addressing digital literacy and data literacy challenges in civic contexts, as highlighted by \cite{mihailidis2018civic} and \cite{bhargava2015beyond}, also informed our approach. Therefore, our design goal is \textit{scaffold data introduction and exploration so that the data and tool are accessible to planners and community members.}

\subsection{\flexiBadge{DG-Flexi} Flexibility and Consistency} From the planners' perspective, decision-making processes happen consistently and repeatedly across planning departments and any democratic space. 
But, planners describe community engagement as a science and an art: \textit{``planning is an art and a science. Science involves technical data and zoning maps, and having the big topics. The art involves the insights and recommendations. From the data we understand problems (insights), how do we address the problems (recommendations)''} (P10). Planners highlighted that need for flexibility as they define the steps of the process. While there is a lot of similarities across projects and engagements, some projects span decades, others months. The phases change depending on the need. Therefore, planners need to be able to reckon with the scale of place, time, complexity, and the various types of stakeholders and goals. Further, they would resist a tool that forces citation and listening in an inauthentic light: \textit{``I don’t want something so rigid where every recommendation has to be tied to an insight because it is always so much more nuanced than that. Having that flexibility and murkiness to allow all the insight to be shared and seen is good, but making sure that nothing we’re doing requires us to constantly tie it to a recommendation''} (P10). Yet, a limitation of some participatory methods can be the bespoke tools created with low to no impact in a broader sense, meeting the need of the specific conditions of a project, but falling short of transferability, though key learnings may transfer. Therefore, we strive to \textit{balance between flexibility and consistency in community engagement efforts.}

\subsection{\relateBadge{DG-Relate} Bridging Relationships} 
In civic planning processes, a critical trust gap emerges when community feedback seems to disappear into a void, only for decisions to materialize later with no clear connection to input gathered. 
This pattern fractures relationships between communities and government, with residents perceiving engagement as merely checking a box rather than genuine listening. 
As C20 expressed: \textit{``People are fed up with the blatant disrespect... entities coming into our community disrespectfully.... We want to be at the table.''} 
This sentiment reflects widespread concerns that plans may be predetermined regardless of community voice. 
Meanwhile, planners articulated their own challenge: \textit{``One of our goals in public engagement is to really try to show and demonstrate that we are taking in that feedback and putting it into our approach and our recommendation''} (P14). 
The complexity extends beyond simple feedback loops to interconnected systems like when comments about trees reveal related sanitation issues requiring coordination across agencies. 
These insights led to our design goal: \textit{Enable the formation of meaningful relationships within civic decision-making that support the iterative nature of these processes.}

\section{Voice to Vision System}


The Voice to Vision system bridges community input and engagement outcomes through a structured data architecture and specialized interfaces for community members and planners.
The foundation of the system involves a data structure comprising interconnected collections that organize community input while maintaining adaptability for diverse engagement contexts. 
This architecture supports two complementary platforms: a community-facing platform that makes engagement processes transparent and accessible to residents, and a sensemaking interface that helps planners organize, analyze, and synthesize community feedback into outputs. 

\subsection{Implemented Data Structure}



At the core of the Voice to Vision system lies a structured yet adaptable data model comprising multiple interconnected collections. 
Each collection contains a predefined set of data fields with expected types, creating a consistent framework for organizing diverse community input data. 
A comprehensive definition of this data structure is available in the supplementary materials (S9).

Our data structure employs a flexible schema model that accommodates the inherently unstructured nature of community input. 
This design choice deliberately balances consistency (through defined data fields and relationships) with flexibility (via optional fields and JSON-compatible inputs), addressing our design goal of maintaining both structure and adaptability throughout the engagement process~\flexiBadge{DG-Flexi}.

Since our implementation is integrated into an ongoing neighborhood planning process with a majority of community engagement completed, we deployed a subset of our proposed data collections, which are displayed in Table~\ref{tab:data_collections}.
This data structure provides the foundation for the platform components described in subsequent sections, enabling both consistent organization and flexible exploration of community input.

\begin{table*}[t]
    \centering
    \caption{Key data collections in the Voice to Vision system architecture.}
    \begin{tabular}{@{}p{0.12\linewidth}|p{0.83\linewidth}}
        \toprule
        \textbf{Collection} & \textbf{Description}\\
        \midrule
        Project&Stores contextual information about the community engagement process, including phase details (start/end dates, status, descriptions). These phases serve as critical metadata for other collections, supporting the documentation of iterative, multi-phase engagement processes~\relateBadge{DG-Relate}.\\
        \midrule
        Events&Captures diverse feedback-gathering mechanisms ranging from surveys to community workshops and tabling events, including descriptive information for each engagement activity.\\
        \midrule
        Voices&Records community input with associated metadata connecting voices to events, topics, and outputs, forming the central repository of community feedback that drives subsequent analysis.\\
        \midrule
        Sub-Geographies&Contains information about distinct areas within the broader study region, particularly relevant for larger planning initiatives that span multiple neighborhoods.\\
        \midrule
        Topics&Maintains information about overarching themes used to organize and make sense of community input. These topics may be predefined or emerge organically through the sensemaking process.\\
        \midrule
        Outputs&Stores the synthesized products of the engagement process, categorized as insights (key findings), goals (high-level visions), and recommendations (specific strategies). Beyond storing descriptive information and voice citations, this collection captures connections between different outputs~\relateBadge{DG-Relate}.\\
        \bottomrule
    \end{tabular}
    \label{tab:data_collections}
\end{table*}


\subsection{Key Platform Components}


\subsubsection{Project Context}

\begin{figure*}[ht]
\includegraphics[width=\linewidth]{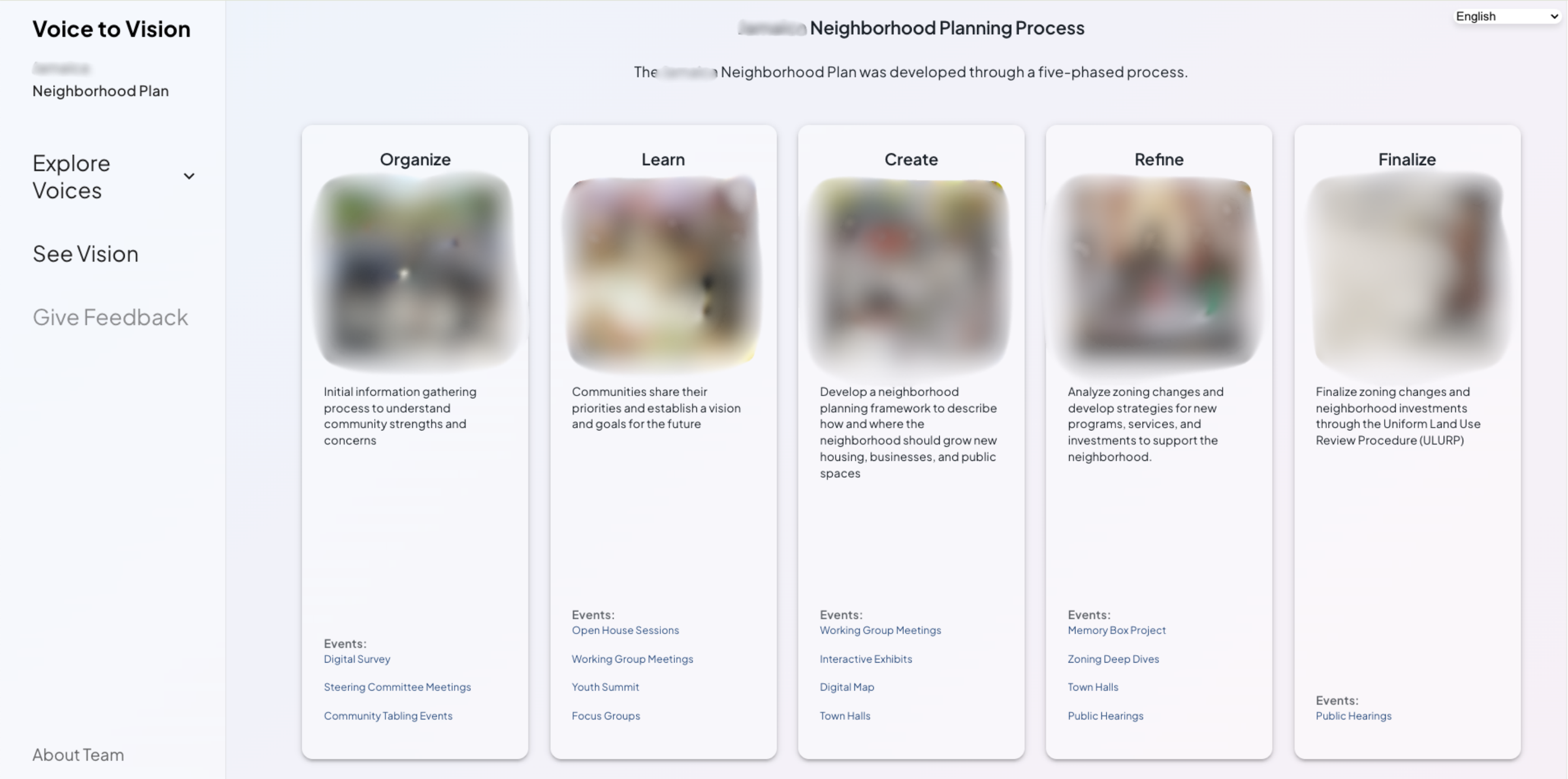}
\caption{The Voice to Vision landing page provides project context and overview to community members, serving as their first interaction with the community-facing platform. The landing page introduces the planning process goals and structure, orienting users before they explore specific community feedback and outputs. We blurred parts of the figure to maintain anonymity.\vspace{-0.3cm}}
\label{fig:landing_page}
\Description{TODO}
\end{figure*}

The project context component serves as an orientation mechanism for both community members and decision-makers, providing a comprehensive overview of the community engagement effort. 
In the community-facing platform, it functions as the landing page, offering users a high-level introduction to the planning process and its objectives (Figure~\ref{fig:landing_page}). 
The project context draws from two primary data collections: the Project collection, which provides core contextual information and phase details; and the Events collection, which supplies event names organized by phase (Table~\ref{tab:data_collections}).
%
By providing clear onboarding information, the component promotes accessibility for newcomers in either stakeholder group~\scaffoldBadge{DG-Scaffold}. 
Drawing from a shared data structure ensures that both planners and community members view identical project goals and overviews, fostering shared understanding.

\subsubsection{Voices}

\begin{figure*}[ht]
\centering
\begin{subfigure}[t]{0.32\linewidth}
    \centering
    \includegraphics[width=\linewidth,height=8cm,keepaspectratio]{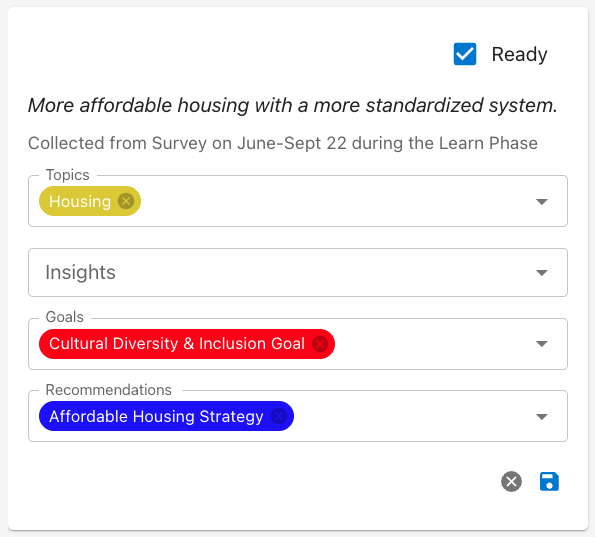}
    \caption{Voice card from the sensemaking interface showing individual community feedback with editable metadata fields. 
    }
    \label{fig:voice_planner}
\end{subfigure}
\hfill
\begin{subfigure}[t]{0.32\linewidth}
    \centering
    \includegraphics[width=\linewidth,height=6.5cm,keepaspectratio]{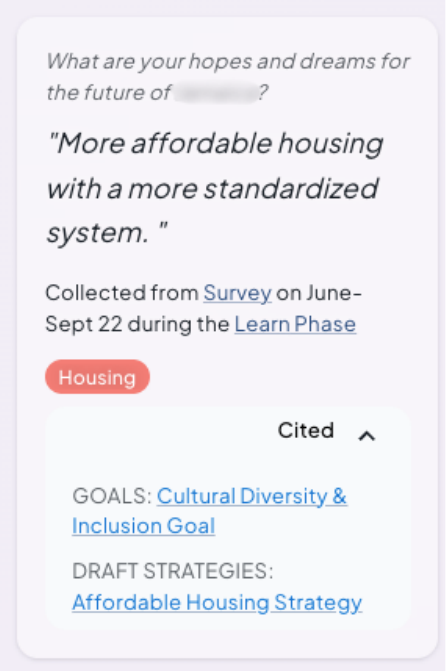}
    \caption{Voice card from the community-facing platform in a read-only format. 
    }
    \label{fig:voice_community}
\end{subfigure}
\hfill
\begin{subfigure}[t]{0.32\linewidth}
    \centering
    \includegraphics[width=\linewidth,height=6.5cm,keepaspectratio]{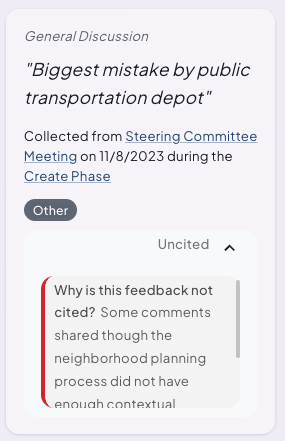}
    \caption{Voice card from the community-facing platform that was uncited, or not tied to specific outputs. 
    }
    \label{fig:uncited_voice}
\end{subfigure}

\caption{Voice cards across both platforms. The sensemaking interface (a) provides editing capabilities for planners to categorize and connect feedback to planning outputs, while the community-facing platform (b) presents the same content in a read-only format that emphasizes transparency, and (c) displays uncited voices with rationales for why they are not tied to specific outputs.}
\label{fig:voice_cards}
\Description{}
\end{figure*}

Voice cards represent the core unit of community input in our system, making individual pieces of feedback accessible across both platforms.
Figure~\ref{fig:voice_cards} compares the voice cards in each platform, highlighting their similarities and differences. 
By surfacing individual contributions in a structured yet rich format, these components make community input visible to both planners and community members~\transpBadge{DG-Transp}.
Both platforms draw from a shared Voices collection, displaying consistent metadata fields including event source, project phase, tagged topics, and relevant outputs (Table~\ref{tab:data_collections}).
A key distinction is that community-facing voice cards are read-only, while planners can edit certain metadata fields (i.e. topics and outputs) within the sensemaking interface during the analysis process.
To populate the voice cards, the system integrates data from multiple collections: Voices, Events, Topics, and Outputs (Table~\ref{tab:data_collections}). 
This integration provides contextual framing for each piece of community input. 
Moreover, voice cards incorporate hyperlinked references to these related elements, allowing users to view dialog windows with descriptions of associated events, topics, or outputs~\relateBadge{DG-Relate}.

The voice card design includes several specialized elements that enhance transparency in the decision-making process. 
A collapsible ``cited'' accordion displays connections between individual feedback and planning outcomes, while also showing when feedback is ``uncited'', or not directly linked to specific outputs. 
For uncited voices, the community-facing platform provides rationales (e.g. insufficient context, outside project scope) that explain why particular feedback wasn't incorporated, addressing a common frustration in community engagement where input seems to disappear without explanation~\transpBadge{DG-Transp}.
Additionally, approximately 14\% (439 / 3037) of community feedback in our case study consisted of audio recordings, which we preserved through an embedded audio player alongside text transcriptions. 
This multi-modal approach honors the original expression while maintaining accessibility, as hearing someone's voice can humanize feedback in ways that text alone cannot~\mirrorBadge{DG-Mirror}~\cite{schroeder2017humanizing}.

\subsubsection{Outputs}

\begin{figure*}[ht]
\centering
\begin{subfigure}[t]{0.48\linewidth}
    \centering
    \includegraphics[width=\linewidth,height=6cm,keepaspectratio]{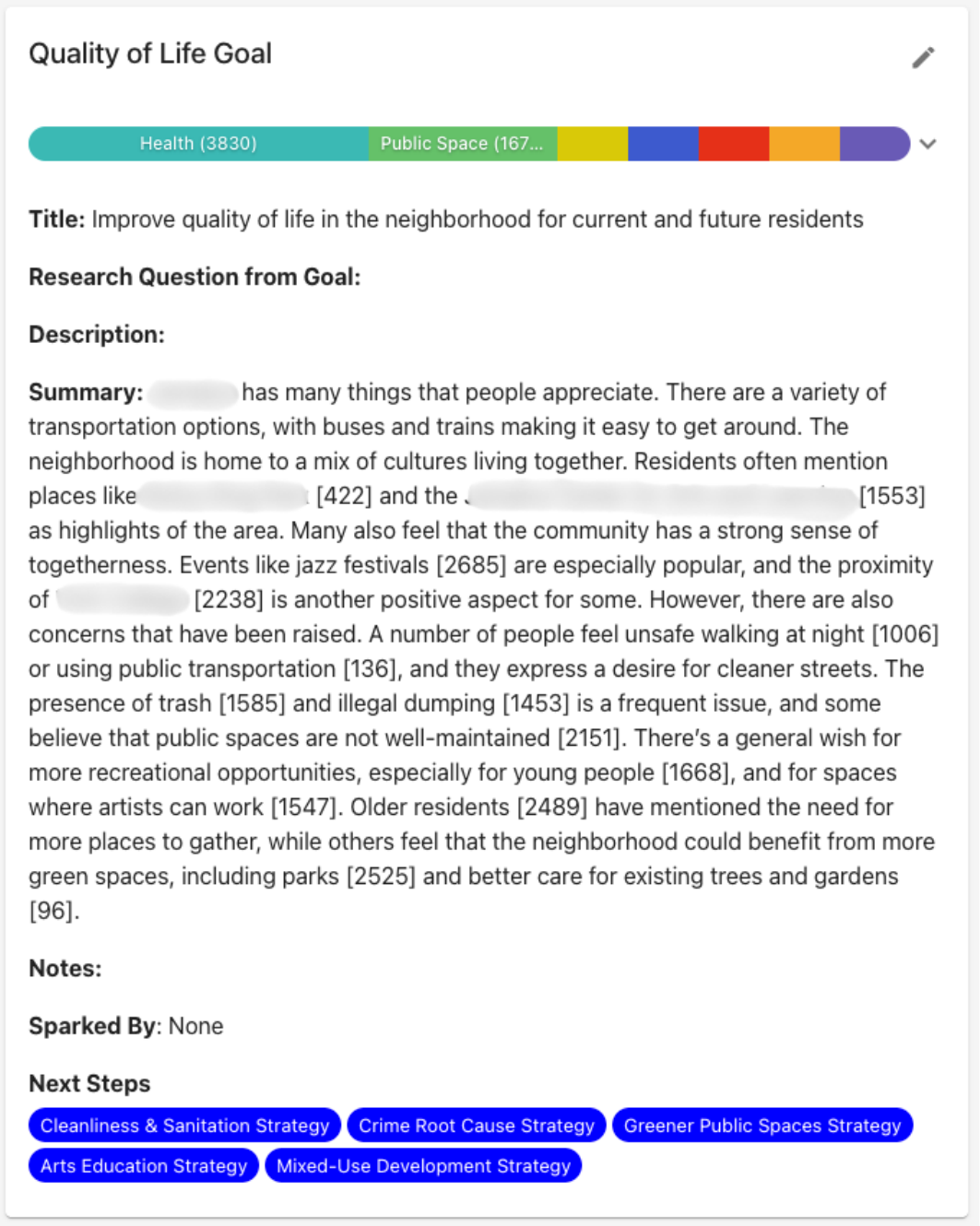}
    \caption{Goal card from the sensemaking interface displaying a high-level planning objective. 
    }
    \label{fig:goal_planner}
\end{subfigure}
\hfill
\begin{subfigure}[t]{0.48\linewidth}
    \centering
    \includegraphics[width=\linewidth,height=6cm,keepaspectratio]{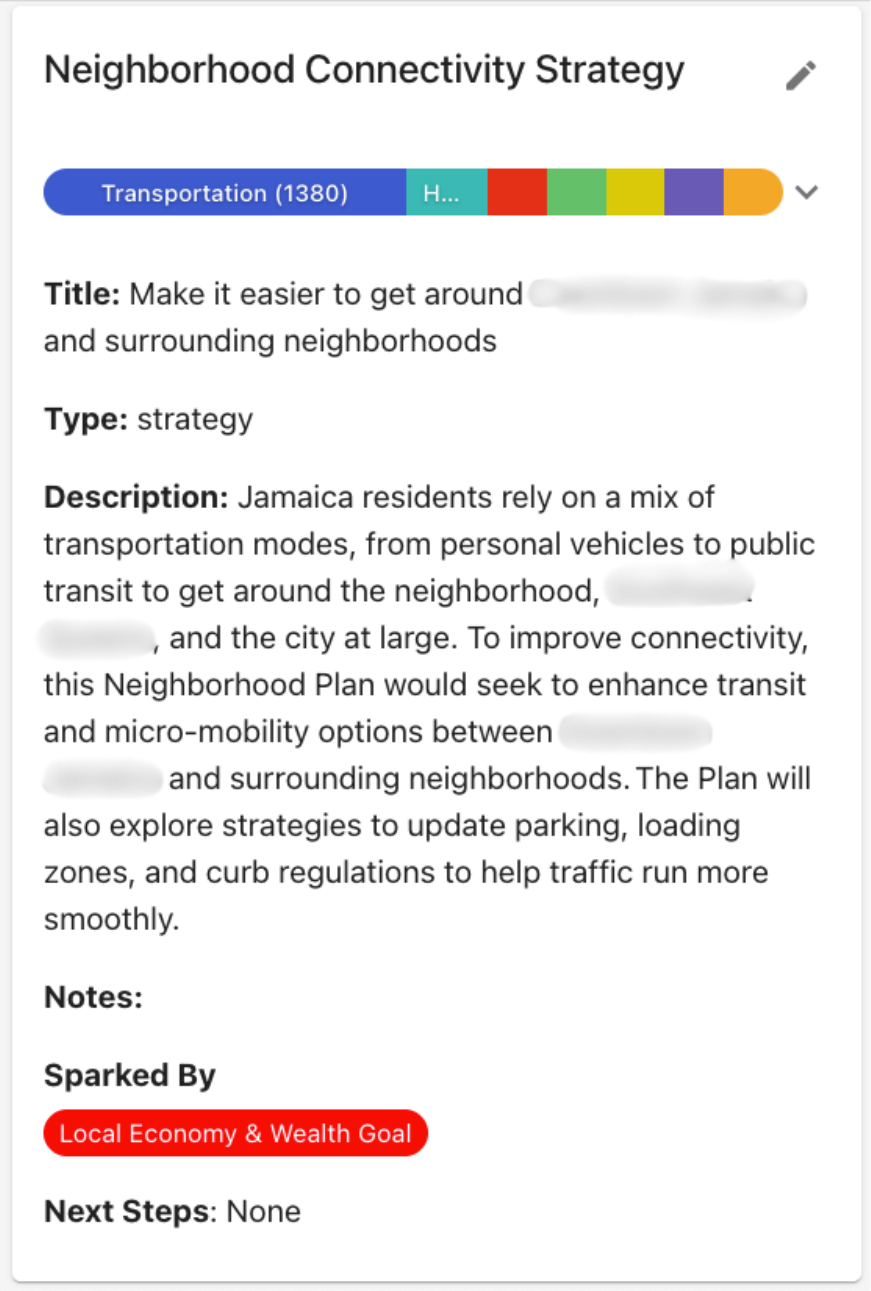}
    \caption{Strategy card from the sensemaking interface showing a specific recommendation. 
    }
    \label{fig:strat_planner}
\end{subfigure}

\vspace{0.4cm}

\begin{subfigure}[t]{0.48\linewidth}
    \centering
    \includegraphics[width=\linewidth,height=6cm,keepaspectratio]{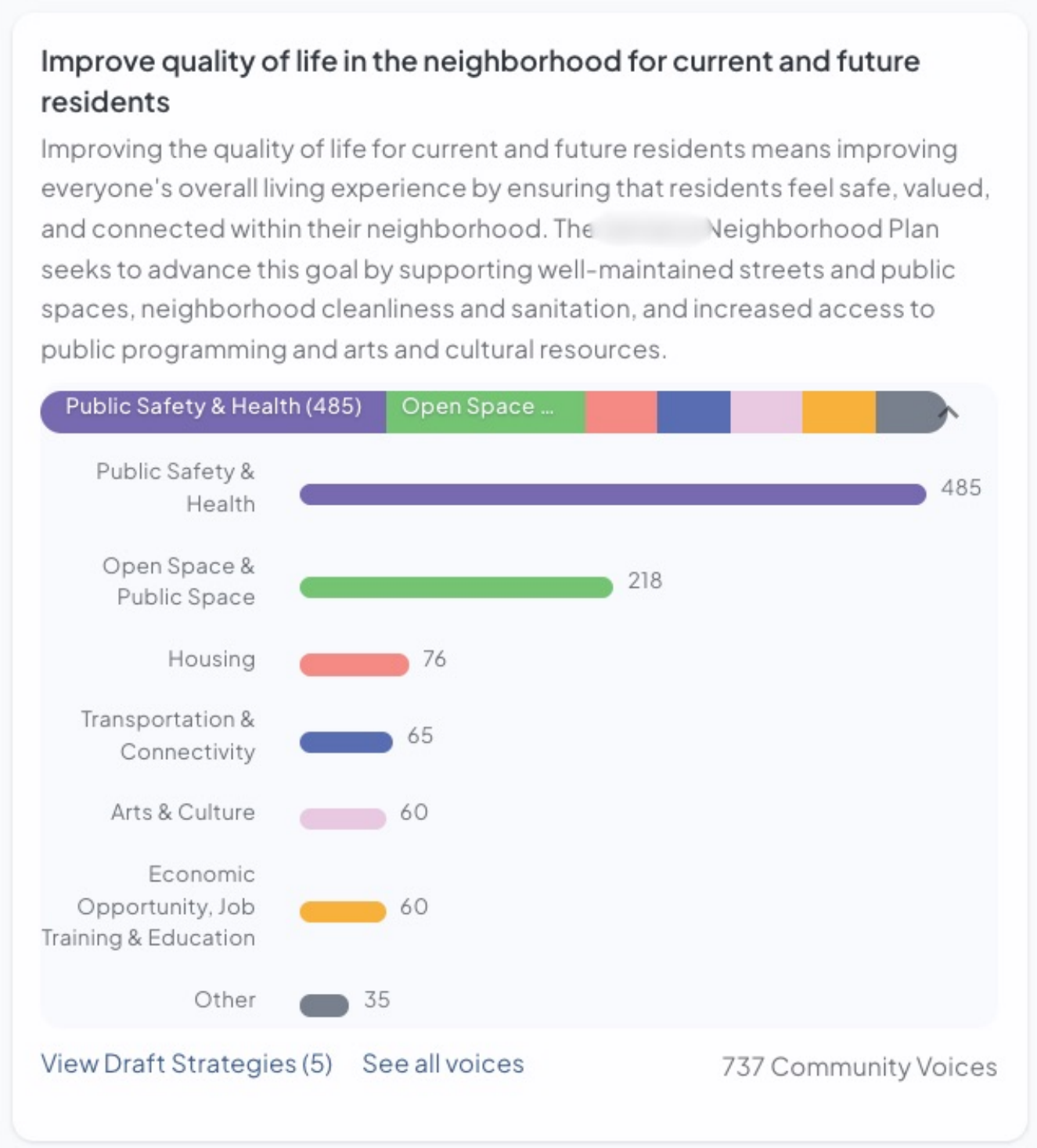}
    \caption{Goal card from the community-facing platform. 
    }
    \label{fig:goal_community}
\end{subfigure}
\hfill
\begin{subfigure}[t]{0.48\linewidth}
    \centering
    \includegraphics[width=\linewidth,height=6cm,keepaspectratio]{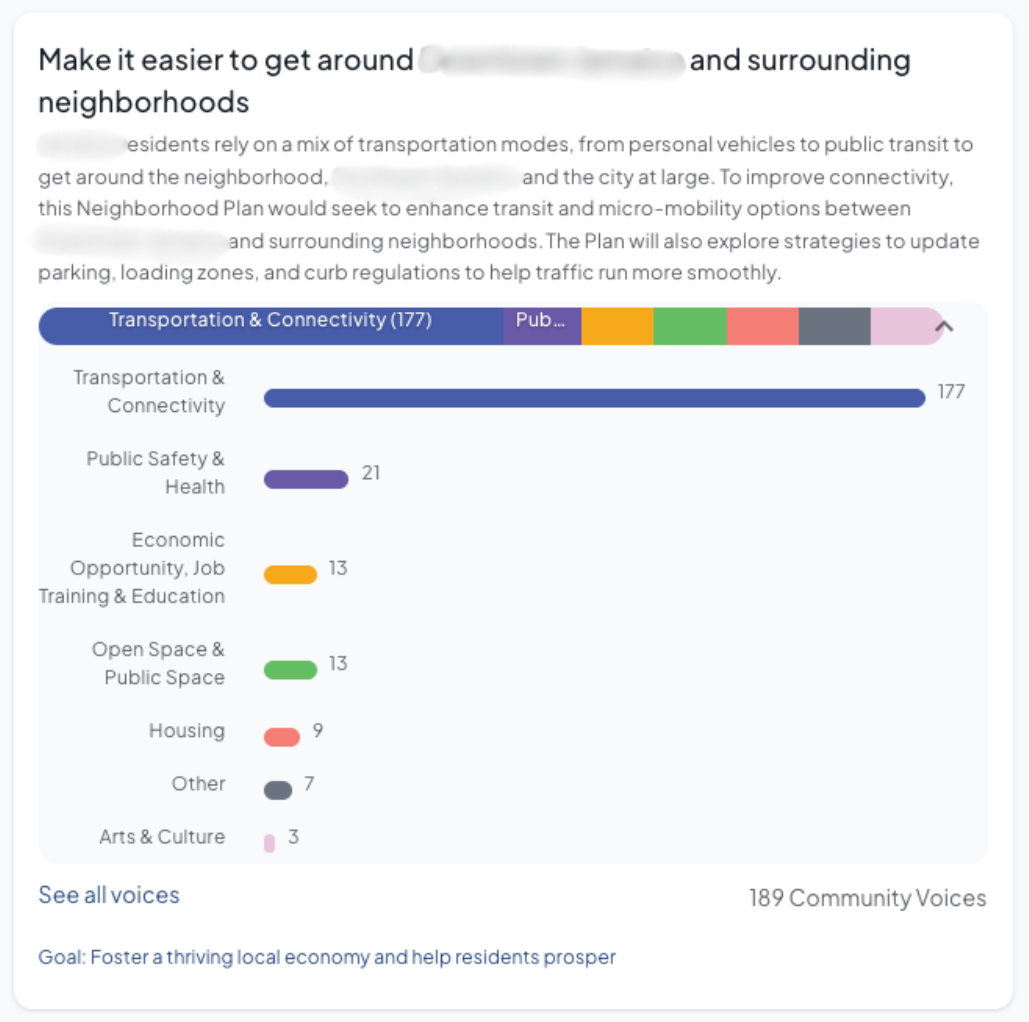}
    \caption{Strategy card from the community-facing platform showing implementation recommendations. 
    }
    \label{fig:strat_community}
\end{subfigure}

\caption{Output cards across both platforms. The top row shows the sensemaking interface's editable output cards used by planners (a, b), while the bottom row shows the community-facing platform's read-only presentations of the same content (c, d). Both implementations maintain consistent metadata fields and interactive visualizations of cited voices by topic.\vspace{-0.5cm}}
\label{fig:output_cards}
\Description{}
\end{figure*}


Engagement outputs are displayed in both platforms through a structured card format as shown in Figure~\ref{fig:output_cards}. 
These cards present key information, including title, description, number of cited voices, and a textual summary of cited voices. 
By maintaining consistent metadata fields across platforms, both stakeholder groups view the same set of outputs, promoting shared understanding.
As with voice cards, the sensemaking interface provides editing capabilities for planners to refine output content while the community-facing cards are read-only.

The system draws from two primary data collections when populating output cards (Table~\ref{tab:data_collections}): the Outputs collection (providing core content) and the Voices collection (supplying metadata on cited voices, particularly their topics). 
This integration enables an interactive visualization feature in the form of a compact stacked bar graph that illustrates the topic distribution of voices cited within each output.
Users can hover over color blocks to identify specific topics or click to expand the visualization for detailed exploration. 
For example, Figure~\ref{fig:goal_community} demonstrates that for the quality of life goal, the majority of cited comments address public safety and health. 
This visualization directly advances our transparency design goal by explicitly showing how community input informed each planning output~\transpBadge{DG-Transp}.

Beyond documenting connections between individual voices and outputs, the system also captures relationships between different types of outputs. 
In the sensemaking interface, these connections are represented through ``sparked by'' and ``next steps'' metadata fields that planners can define. 
The community-facing platform translates these relationships into more intuitive interfaces by displaying goals and strategies side-by-side and enabling users to filter strategies by selecting specific goals to see their associated recommendations. 
This implementation addresses our bridging relationships design goal~\relateBadge{DG-Relate} by making explicit how goals generate specific strategies, providing community members with a clearer understanding of the planning process from findings to recommendations.

\subsubsection{Visualizations}

\begin{figure*}[ht]
\centering
\begin{subfigure}[b]{\linewidth}
    \centering
    \includegraphics[width=\linewidth]{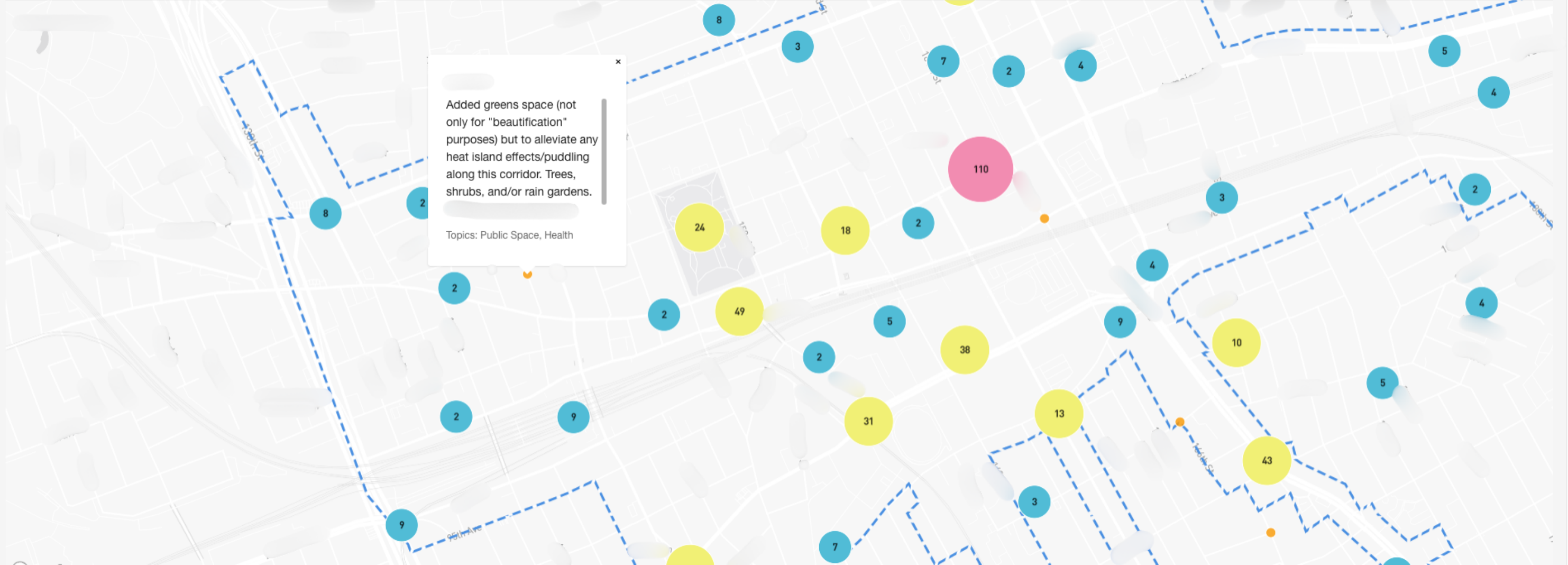}
    \caption{Interactive map visualization from the sensemaking interface showing geotagged community voices. 
    }
    \label{fig:map}
\end{subfigure}

\vspace{0.2cm}

\begin{subfigure}[b]{\linewidth}
    \centering
    \includegraphics[width=\linewidth, height=8cm, keepaspectratio]{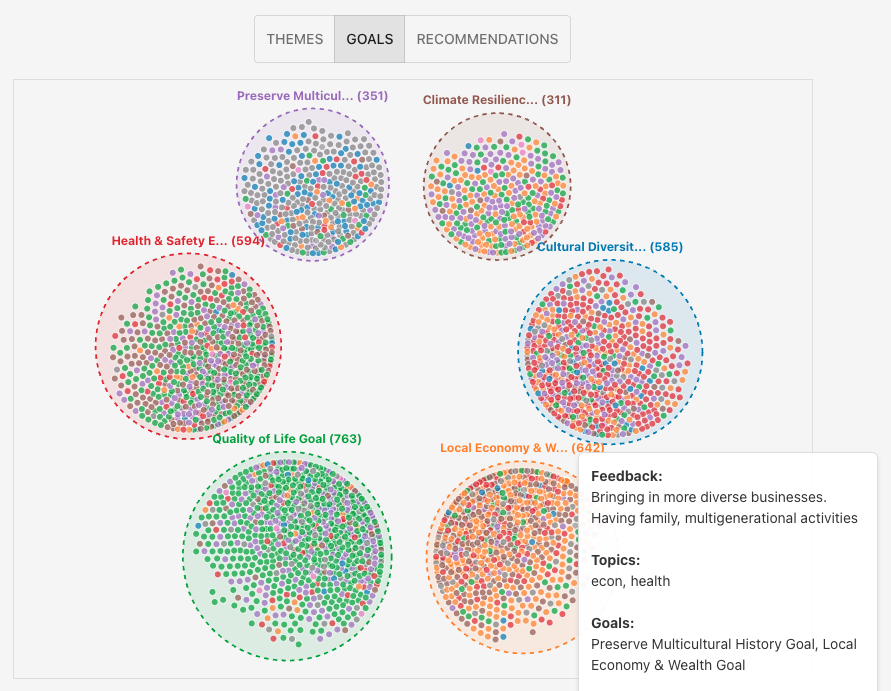}
    \caption{Cluster visualization from the sensemaking interface organizing community voices by goals. Circle size represents the number of associated voices in each category, with individual voices appearing as smaller color-coded points within each cluster.}
    \label{fig:cluster}
\end{subfigure}

\caption{Visualizations in the sensemaking interface. The map visualization (a) reveals spatial relationships while the cluster visualization (b) highlights thematic connections across the voices.}
\label{fig:visualizations}
\Description{}
\end{figure*}


Large community engagement initiatives generate substantial volumes of feedback.
For instance, our case study accumulated over 3,000 pieces of community input. 
While the system's default presentation is a simple list view of comments, this approach quickly becomes unwieldy at scale. 
To address this challenge, we implemented complementary visualization approaches that facilitate more accessible explorations of community feedback~\flexiBadge{DG-Flexi}.

Our visualization components include an interactive map (Figure~\ref{fig:map}) and a dynamic cluster view (Figure~\ref{fig:cluster}), both drawing from multiple data collections (Table~\ref{tab:data_collections}): Voices (providing feedback text, geographic coordinates, and metadata tags), Topics (supplying topic names), and Outputs (connecting feedback to outcomes). 
The interactive map visualization (Figure~\ref{fig:map}) displays geotagged voices as color-coded dots, with each color representing one of the voice's associated topics. 
This component processes location references in community feedback using the Google Maps Address Validation API\footnote{\url{https://developers.google.com/maps/documentation/address-validation}.} to convert textual locations into geographic coordinates.
Upon clicking any point, users can view the complete feedback text alongside relevant metadata, similar to the voice card design. 
To prevent visual overload, we implemented a responsive clustering algorithm that dynamically groups nearby points based on the current zoom level, progressively revealing more granular detail as users zoom into specific neighborhoods.

Complementing the spatial view, the cluster visualization (Figure~\ref{fig:cluster}) provides a high-level thematic organization of community input. 
This component offers three organization schemes: by topic, goal, or recommendation. 
Each scheme generates a circular arrangement where individual circles represent discrete categories (e.g., specific topics), with circle size proportional to the number of associated voices. 
Within each circle, individual voices appear as smaller color-coded points that can be hovered over to reveal their content. 
This visualization is particularly effective for identifying patterns in feedback distribution across different dimensions, and was not included in the community-facing platform due to time constraints during the field deployment period, though we received some initial positive feedback from a few community members.

Both visualizations directly address our bridging relationships design goal~\relateBadge{DG-Relate} by highlighting the connections between different data elements. 
The map visualization reveals spatial relationships between voices, topics, and sub-geographies, while the cluster visualization highlights thematic connections between voices and either topics or outputs. 
Both visualizations support users in recognizing patterns that might remain obscured in text-only presentations, thereby facilitating more nuanced understanding of community priorities.

\subsubsection{Interactions}


While our visualization components offer rich exploratory capabilities, we complemented these with several familiar data interaction features (i.e., filtering, sorting, and searching) designed to provide intuitive ways for users to navigate through large volumes of community input~\scaffoldBadge{DG-Scaffold}.
The filtering system allows users to refine the displayed voices based on metadata from four key data collections (Table~\ref{tab:data_collections}): Events (isolating feedback from specific engagement activities), Sub-Geographies (focusing on particular neighborhoods or areas), Topics (highlighting feedback addressing specific themes), and Outputs (showing voices that informed particular planning outcomes). 
The sorting functionality enables chronological organization of feedback by project phase, and the search capability implements keyword matching, allowing users to quickly locate feedback related to specific terminology.
The three interaction features are implemented consistently across both platforms, giving community members and planners access to the same exploration abilities.
The combination of standard interaction patterns with more sophisticated visualizations creates a layered exploration experience that accommodates diverse user preferences and technical comfort levels. 

\subsection{Implementation Details}

We implemented the Voice to Vision system's data structure using MongoDB\footnote{\url{https://www.mongodb.com/}.}, a NoSQL database that aligns with~\flexiBadge{DG-Flexi} by accommodating both structured fields and unstructured community input. 
Both the community-facing and sensemaking platforms were built using a consistent tech stack: React\footnote{\url{https://react.dev/}.} with React-Query\footnote{\url{https://tanstack.com/query/latest}.} for state management and Material UI\footnote{\url{https://mui.com/}.} for interface components on the frontend, with a Flask web framework\footnote{\url{https://flask.palletsprojects.com/en/2.2.x/}.} serving as the backend. 
The systems were containerized using Docker\footnote{\url{https://www.docker.com/}.} and deployed through Amazon Web Services\footnote{\url{https://aws.amazon.com/}.}.
We added additional features to the community-facing platform to ensure it was mobile-responsive and multilingual (integrated with Google Translate Widget) to broaden participation~\cite{toukola_digital_2022, sarangapani_virtualculturalcollaboration_2016}.

\section{Evaluation Methodology}

Our evaluation approach combines multiple methods to assess both components of our system.
We evaluated the internal sensemaking interface through structured sessions with seven urban planners, 
providing valuable early indicators of effectiveness despite the relatively small sample size. 
For the community-facing platform, we conducted a public field deployment in a diverse urban neighborhood as part of an actual planning process, tracking usage analytics and gathering qualitative feedback from 17 community members. 
This approach suitably reflects the distinct purposes of each platform: the sensemaking interface as an internal professional tool evaluated through expert assessment, and the community platform as a public interface evaluated through a real-world deployment. 

\subsection{Planner User Studies}

While our iterative design process yielded numerous insights into the sensemaking process, we conducted a separate assessment of the sensemaking interface to evaluate a comprehensive prototype that was an output of the design process. 
The evaluation consisted of an in-person 2-hour workshop with three planners from our design cohort and three 30-minute individual interviews with planners outside our cohort. 
While we acknowledge the limitations of this relatively small sample size, these assessments provide valuable early indicators of the platform's effectiveness and usability.
We emphasize that we are currently collecting feedback from additional planners to strengthen these preliminary findings.

\begin{table*}[t]
    \scriptsize
    \centering
    \caption{Information about each participant in the planner user studies.
    ``Role'' indicates the participant's professional position. 
    ``Exp (yrs)'' shows the participant's years of experience in urban planning.
    ``Specialty'' describes the participant's area of planning expertise. 
    ``Session Format'' indicates whether the participant took part in the workshop or individual interview session. 
    Participant P2 (*) is a co-author of this paper.}
    \begin{tabular}{@{}p{0.05\linewidth}|p{0.2\linewidth}|p{0.1\linewidth}|p{0.35\linewidth}|p{0.15\linewidth}}
        \toprule
        \textbf{ID}&\textbf{Role}&\textbf{Exp (yrs)}&\textbf{Specialty}&\textbf{Session Format}\\
        \midrule
        P1&Senior Lead&12&Urban Design&Workshop\\
        P2 (*)&Director&20&Civic Engagement&Workshop\\
        P3&Senior Planner&7&Housing&Workshop\\
        P4&Associate Planner&8&Transportation&Interview\\
        P5&Senior Planner&8&Neighborhood Planning&Interview\\
        P6&Project Manager&4&Environmental Review&Interview\\
        P7&Senior Planner&15&Neighborhood Planning&Interview\\
        \bottomrule
    \end{tabular}
    \label{tab:plannerparticipants}
\end{table*}

\subsubsection{Participants}

We conducted an initial evaluation of the sensemaking platform with seven urban planners (P1-P7). 
Three participants (P1-P3) were part of our design cohort and participated in an in-person workshop, while four additional planners (P4-P7) outside our design cohort participated in individual interviews. 
Participant P2 is also a co-author on this paper, bringing both researcher and practitioner perspectives to our evaluation. 
Table~\ref{tab:plannerparticipants} provides details on participants' planning backgrounds.
Participation was voluntary and participants were not compensated for participating in the evaluation.

\subsubsection{Procedure}

For the in-person workshop, we first provided participants (P1-P3) approximately 20 minutes to explore the sensemaking platform independently on their laptops. 
This allowed them to develop initial impressions without researcher guidance. 
We then systematically reviewed each component of the platform, soliciting feedback through guided discussion. 
We asked follow-up questions like, \textit{Which visualizations were effective? Which were less so?} to facilitate rich dialogue about the prototype's strengths and weaknesses.

For the one-on-one interviews with participants P4 to P7, we created a condensed version of the workshop protocol to account for the shorter interview times. 
These 30-minute sessions were conducted remotely over Zoom\footnote{\url{https://www.zoom.com/}}. 
Participants were given access to the platform and asked to think aloud while exploring its features.
We asked follow-up questions to elicit insights about strengths and weaknesses. 
Each interview concluded with reflection questions such as, \textit{How could this prototype fit into your current planning work?} The complete list of questions asked in the workshop and interviews are available in the supplementary materials (S10-S11).

\subsubsection{Data Collection and Analysis}

We recorded and transcribed both the workshop and interviews to facilitate detailed analysis. 
The transcripts were segmented into units of analysis based on paragraphs and discussion topics.
We analyzed the transcript data using affinity diagramming.
The second author conducted an initial clustering of notes, which was subsequently reviewed by the first, third, and sixth authors, resulting in three clustering iterations.
%

\subsection{Field Deployment}

To evaluate our community-facing platform under real-world conditions, we deployed it as part of an ongoing neighborhood planning process, tracking usage patterns and collecting feedback from actual community members interacting with the system in its intended context.

\subsubsection{Deployment Neighborhood}
Our community-facing platform was deployed in a diverse urban neighborhood in a major U.S. metropolitan area. 
The area's approximately 256,278 residents represent a rich multicultural heritage, with significant Black, Hispanic and Asian populations. Approximately 44\% of the area's residents are foreign born and speak a wide range of languages including Spanish, Bengali, Urdu, Haitian Creole, and other languages. Housing stock varies from higher-density apartments in the central district to single and two-family homes in surrounding areas, with home ownership rates slightly above the citywide average.

The neighborhood serves as a significant transit hub with multiple subway lines, regional rail connections, extensive bus service, and major driving corridors. Many residents rely on a combination of public transit for their 48-minute average daily commute and personal vehicles. 

The neighborhood functions as an important employment center with approximately 24,000 jobs concentrated in retail, education, healthcare, transportation, and hospitality sectors. 
Community assets include 14 public parks totaling 29 acres, approximately 3,700 street trees, multiple cultural institutions, and an active network of community-based organizations. 
While 79\% of residents report good to excellent health, the area faces environmental challenges including higher surface temperatures and elevated childhood asthma rates compared to surrounding areas. 
This combination of transit connectivity, cultural diversity, and mixed residential-commercial character makes the neighborhood an ideal testing ground for civic technology aimed at inclusive community planning.

\subsubsection{Deployment Details}
The community-facing platform was launched in March 2025 as part of the neighborhood's planning process.
City planners formally introduced the platform through a press release that accompanied the draft neighborhood plan when it entered the formal public review process, establishing its role as an official digital complement to the planning process. 
To promote community awareness and participation, we employed multiple outreach strategies, including social media posts on the local city council members' and city planning accounts, physical posters displayed at the local library, general email announcements through the city's planning newsletter, and targeted emails to community organizations and previous planning meeting attendees. 
The platform remains active as of this writing and will continue to operate until the neighborhood planning process concludes at the end of 2025, providing a nine-month window for community engagement and data collection. 

\subsubsection{Data Collection and Analysis}

We tracked platform usage through anonymous session IDs provided by browser cookies and maintained through Flask-Session and MongoDB.
Since browser cookies periodically expire, session IDs do not necessarily represent unique users, an important limitation of our tracking approach. 
The platform collected two types of interaction data: feature usage events (capturing specific user actions) and heartbeat data at 5-second intervals (recording page location, device type, and language settings). 
We also implemented a dedicated feedback page with both closed and open-ended questions.
However, we only received one response during this early deployment phase. 
For analysis, we quantitatively examined the event tracking and heartbeat data from March to May 2025 to calculate metrics including feature usage frequency, navigation patterns, and session duration. 
To reduce the impact of outliers, we filtered the dataset by removing five session IDs representing the top 5\% of users by session count, resulting in data from 84 unique session IDs. 
Analysis of usage patterns revealed an average session duration of 7.91 (+/- 15.51) minutes, suggesting substantial engagement with the platform despite considerable variance in visit length. 
While we attempted to filter out research team and user study visits from our analytics, some research-related interactions may remain in the dataset, potentially affecting our usage metrics.

\subsection{Community User Studies}

To obtain richer qualitative insights into how the community-facing platform addresses our design goals, we conducted 15 user studies with 17 community members.

\begin{table*}[t]
    \scriptsize
    \centering
    \caption{Information about each participant in the community user studies.
    ``ID'' refers to the participant identifier.
    ``Age'' indicates the participant's age range in years. 
    ``Race/Eth'' refers to the participant's self-identified racial or ethnic background. 
    ``Res (yrs)'' shows how long the participant has lived in the neighborhood. 
    ``Imm'' indicates immigrant status (Y=Yes, N=No). 
    ``Tech'' represents the participant's self-reported comfort level with technology (Low, Med, High). 
    ``Format'' describes whether the session was conducted remotely (R) or in-person (IP).  
    ``Involvement'' summarizes the participant's self-reported level of engagement with community activities. Participants C14 to C17 did not complete the pre-survey, so their fields are blank.}
    
    \begin{tabular}{@{}p{0.06\linewidth}|p{0.04\linewidth}|p{0.15\linewidth}|p{0.07\linewidth}|p{0.07\linewidth}|p{0.04\linewidth}|p{0.06\linewidth}|p{0.31\linewidth}}
        \toprule
        \textbf{ID}&\textbf{Age}&\textbf{Race/Eth}&\textbf{Res (yrs)}&\textbf{Imm}&\textbf{Tech}&\textbf{Format}&\textbf{Involvement}\\
        \midrule
        C1&25-34&Asian&1-5&Y&High&IP&No involvement in community activities and planning\\
        C2&35-44&Black or African American&20+ years&N&High&IP&Very involved in community activities and planning\\
        C3&18-24&South Asian&11-20&Prefer not to answer&High&IP&Moderate involvement in community activities, no involvement in planning\\
        C4&45-54&Hispanic or Latino&All my life&N&Med&R&Little involvement in community activities, no involvement in planning\\
        C5&25-34&Black or African American, Caribbean&All my life&N&High&R&Moderate involvement in community activities and planning\\
        C6&35-44&Black or African American&20+ years&N&High&R&Very involved in community activities, moderate involvement in planning\\
        C7&65-74&Black or African American&20+ years&N&High&R&Very involved in community activities, moderate involvement in planning\\
        C8&18-24&Black or African American&6-10&N&High&R&Very involved in community activities and planning\\
        C9&35-44&Hispanic or Latino, White or Caucasian&11-20&N&High&R&Very involved in community activities, little involvement in planning\\
        C10&55-64&Black or African American&All my life&N&High&R&Very involved in community activities and planning\\
        C11&55-64&Prefer not to answer&11-20&Y&High&R&No involvement in community activities and planning\\
        C12&35-44&Black or African American&All my life&N&High&R&Very involved in community activities, moderate involvement in planning\\
        C13&25-34&South Asian&1-5&Y&High&R&No involvement in community activities and planning\\
        C14-16&---&---&---&---&---&IP&---\\
        C17&---&---&---&---&---&R&---\\
        \bottomrule
    \end{tabular}
    \label{tab:communityparticipants}
    \vspace{-0.4cm}
\end{table*}

\subsubsection{Participants}

We recruited 17 community members (C1-C17) to participate in our user studies. 
Participants represented a diverse cross-section of the neighborhood's demographic profile, with varied levels of prior engagement with community planning processes. 
Table~\ref{tab:communityparticipants} summarizes participant demographics, including age range, racial background, length of residence, technology comfort level, and prior involvement in local planning activities.
%
Participants' technology comfort levels were mostly high.
Pre-survey responses revealed a diverse range of perspectives regarding the planning process. 
While participants largely believed they could contribute to shaping neighborhood plans (9 agree, 4 neutral, 0 disagree), there was more variation in whether they felt their thoughts had been heard and reflected in the plan (6 agree, 4 neutral, 3 disagree). 
Trust in city planners showed a similar distribution (7 agree, 3 neutral, 3 disagree). 
Most participants indicated they understood how community feedback informed the plan (9 agree, 2 neutral, 2 disagree), and nearly all expressed willingness to participate in future planning efforts (12 agree, 1 neutral, 0 disagree). 
This distribution suggests our sample included both highly engaged community members and those more skeptical of planning processes.

We conducted four sessions in person and ten sessions remotely. 
In-person sessions were held in two locations: a local library where we invited community residents to provide feedback on-site, and the office of a city council member with jurisdiction over the planning area. 
The council office session included three participants (C14, C15, C16) who worked for the council member.
All participants except the councilor's staff members and C17, who was connected with the planners, filled out the pre-survey and received \$30 Visa gift cards as compensation for their time.
We did not offer compensation to the councilor's staff and C17 to avoid any ethical concerns due to them being involved in the project.
As such, their involvement in the evaluation remains completely voluntary.

\subsubsection{Procedure}

User study sessions lasted 30 to 60 minutes and followed a three-part structure. 
First, participants completed a pre-survey with demographic questions and six Likert-scale items measuring initial impressions on key outcome variables related to civic engagement (e.g., \textit{I can contribute to the shaping of neighborhood plans in my neighborhood}). These questions assessed baseline perceptions about planning process legitimacy and feelings of being heard.

Second, we conducted a guided walkthrough of the platform. 
We first provided a high-level overview of each page, then allowed participants to explore areas of interest while thinking aloud about their observations and questions.
We prompted participants to share their thoughts and also highlighted features they might have overlooked.

Finally, we conducted a post-walkthrough interview with questions that paralleled our pre-survey outcome measures, including \textit{What parts of the tool worked/didn't work for you?} and \textit{How much do you think your voice or voices like yours influence decision making?} The complete pre-survey and post-interview protocols are available in the supplementary materials (S12-S13).

\subsubsection{Data Collection and Analysis}

Demographic data collected from the pre-survey was used to contextualize interview findings and is presented in Table~\ref{tab:communityparticipants}. 
We analyzed the Likert-scale responses by calculating descriptive statistics for each item and visualizing response distributions to identify patterns in participants' perceptions of civic engagement processes.

All user study sessions were recorded and transcribed verbatim. 
We analyzed the transcripts using inductive thematic analysis~\cite{braun2006using}, which involves iterative coding to identify emergent patterns in the data. 
Acknowledging that our pre-survey outcome measures may have influenced our analytical lens, we took deliberate steps to engage with these potential biases throughout our analysis process.
First, we segmented the transcripts into paragraphs as units of analysis, then the first two authors conducted open coding on a random sample of 4 interviews (27\% of the data).
The research team clustered the open codes into categories, refining them into a codebook with 29 thematic codes (see S14 in supplementary materials).
From there, the first three authors collaboratively applied the codebook to 3 interviews to refine the definitions and address possible ambiguities.
After updating the codebook, the first three authors independently coded the remaining 12 interviews using a spreadsheet application, where two coders were randomly assigned to each interview. 
All disagreements were resolved through periodic discussions during the coding process. 
The final inter-coder reliability was 0.645, measured using Cohen’s Kappa~\cite{cohen1960coefficient} and indicating substantial agreement.
The entire coding process took around two weeks to complete with around 35 
hours of independent coding and 10 
hours of live discussion.
After coding was complete, the research team had two discussions to condense the codebook into 5 key themes, presented in Table~\ref{tab:themes}.

\section{Results}


Our evaluation revealed five interconnected dimensions that shape how community members and planners engage with the Voice to Vision system. 
As one community member articulated, when community voices are shared, \emph{``it's important that the information and data that was gathered through these sessions from the people, is heard by the people who are creating this plan... if those things are heard then we can get closer to developing a plan that actually helps the community''} (C7). 
This statement captures the central tension our findings explore: the challenge of transforming collected feedback into meaningful civic action while maintaining transparency and trust. 
Below, we examine how our system addresses these challenges through supporting connection and mirroring between community members, establishing transparency that enhances legitimacy, providing accessibility through scaffolded interfaces, balancing flexibility with consistency across planning contexts, and revealing relationships between data elements and stakeholders. 

\begin{table*}[t]
  \caption{Themes from thematic analysis of interviews with community members. Major themes include specific sub-themes where applicable. Each row includes a description and associated codes.}
  \label{tab:themes}
  \begin{tabular}{@{}p{0.14\linewidth}|p{0.12\linewidth}p{0.31\linewidth}p{0.36\linewidth}@{}}
    \toprule
    \textbf{Themes} & \textbf{Sub-Themes} & \textbf{Codes} & \textbf{Descriptions}\\
    \midrule
    Community Connection & --- & personal experience/opinion, specific interest, different than others, alignment with others, community assets, connection with others, assumptions about others, interexpansion & How community members relate to each other, share experiences, and form collective identity through recognizing similarities and differences in perspectives\\
    \midrule
    \multirow{3}{*}{Transparency} & Fairness & diversity, access, representation, interest in participation & The extent to which the planning process includes diverse voices and provides equitable access to participation opportunities\\
    \cmidrule{2-4}
    & Legitimacy & legitimacy, feeling heard & Community perceptions about whether their input is genuinely considered by decision-makers\\
    \cmidrule{2-4}
    & Actionability & actionability, having an impact & Whether community input translates into concrete actions in the planning process\\
    \midrule
    \multirow{2}{*}{Interface} & Platform Design & confusion/questions, target audience, design feedback, satisfaction, platform accessibility, big picture view & User experience of the system including clarity, accessibility, and overall presentation of information\\
    \cmidrule{2-4}
    & User Behaviors & not exploration, exploration & Patterns of interaction with the platform, including exploration tendencies\\
    \midrule
    Iteration & --- & intraexpansion, time, provenance & How the platform supports ongoing refinement of ideas and feedback over time\\
    \midrule
    Understand & --- & learning, context & How the platform facilitates learning about the planning process and contextualizing community input\\
    \bottomrule
  \end{tabular}
\end{table*}

\subsection{How Reflecting Voices Builds Community Connection}

Multiple themes arose regarding the platform's goal of increasing the visibility of voices within the community~\mirrorBadge{DG-Mirror}. The first was an increased perception of voices and groups of opinions, with 70\% of users viewing the voice cards. Participant C13 mentioned how seeing community members' voices \emph{``resonates with you like you’re not alone, and they’re all tons of people that have the same concerns, and are striving towards, or hoping towards some changes being made to their community or the situation''}. Another participant, C12, was able to make better sense of the differences in opinion they had with others and reflect on various ways people approach change in the neighborhood: \emph{``They’re talking about nostalgic moments. Meanwhile, people’s lives are at stake… Like, this is crazy. It wouldn’t be bad to show like what the nostalgic moments versus the reality, that’s not a bad space to take on, and I think it’s important to really speak to that, because people, they hold on to their ideas. But sometimes they don’t necessarily understand how they affect others''}. This insight relates to the large standard deviation in the number of voice cards users viewed: +/- 219.50. Participants whose views did not resonate with those on the platform may not have explored additional voices. However, participants did feel the map visualization helped place different groups of voices within the neighborhood and enhance understanding of what areas are facing specific challenges. This was also shown through the widespread use of geographic filters in the platform, with those being the most commonly used. 

We also observed how participants perceived the makeup of the community within and outside the platform. Many recognized the diversity of the neighborhood’s residents among the voices, with C10 commenting that neighborhood \emph{``residents are not a monolith, right? And it’s not just based on race, because oftentimes people... speak of it only in terms of racial or ethnic diversity. But there’s, there’s age diversity. There’s income diversity right here. When I see this, I can also tell you that this sounds like a response from newer residents''}. Recognizing the makeup of the voices once again illuminated the different groups within the neighborhood and the alignment or differences between them. For some participants, seeing and hearing community voices allowed them to identify valuable community assets, such as community centers, restaurants, or religious centers. Participant C8 especially felt a sense of gratitude for the role assets played in uplifting neighborhood residents: \emph{``These religious centers help a lot with the community. They’re good business here, too. But if it wasn’t for some of these churches, some of these places wouldn’t have food banks and things like that''}.

Those who perceived discrepancies between the voices in the platform and real life challenged the planning process and called for a more representative manner of data collection. C9 mentioned that \emph{``until we help the ones that need the most help, we won’t be able to flourish as a part of the city. So, I would really hope that those voices are the ones that are heard... How can we help them?''}

\subsection{Limitations of Transparency for Legitimacy without Actionability}
Regarding transparency~\transpBadge{DG-Transp}, openness, and rigor in the process did improve the sense of legitimacy.
However, transparency can be quickly undermined if the points of participation feel inaccessible or exclusive. For example, C12 explains \textit{``I don't necessarily think we get heard because we are... we have yet to even have a space in Jamaica,''} or if not exclusion, uncertainty as C3 states: \textit{``A lot of people,they just don't know the right person to reach out to, to get their voice heard, so I think that's the biggest problem.''} 

Further, citations of voices in outputs proved valuable, C3 observes: \textit{``So, that's definitely helpful. It's, like, you have to kind of compare the ratio of the cited to uncited to see how much of the opinions are being used in public meetings or official hearings.''} 
Planners acknowledge the importance of sharing uncited voices with P6 stating: \emph{``[It is] interesting to see how the other tags manifest and important that we put it out there. It's important for people to see that people have that opinion.''}
However, citations were not explored widely across community members. According to the usage logs, citations were expanded by 35.7\% of users, with a mean event of 1.45 +/1 2.73 per user. The proportion of uncited voice cards expanded tends to be a bit less than the proportion of uncited voices in the data (13.11\% vs 15.48\%), suggesting that while conceptually this feature offers value for legitimacy, it wasn't investigated deeply by users.

One of the most apparent challenges to legitimacy was the actionability of the outputs. C7 explains: \textit{``[These people are] saying what they either did in the past or what they want to do. But I'm not hearing a real, concrete strategy. How do we get there? How do we create these things?... If you're doing a strategy, you're saying, well, we're going to do A, B, C, and D, and anticipate an outcome that's going to reach the goal. I'm not seeing that. It's very vague to me.''} This sentiment of vagueness was repeated broadly among seven community participants. In the overall interaction data, 46.4\% of participants viewed output cards, and 34.5\% dove into the cards with 2.45 +/- 4.91 average deep dive clicks per user. Of those participants, 56\% filtered the outputs in some way, often by goal. From this, we see an interest in exploring the outputs from users, though if the actual contents are not concrete or actionable enough, patience and trust can diminish, leaving some like C7 with a sense of uncertainty: 
\textit{``Well, I think the community came out and voiced what they felt. And again the answer is, gonna be, we're gonna see if their voices were taken into consideration with the end product.''}


Finally, C13 highlighted an interest in not only seeing the actions and promises of the city, but also having the chance to be pointed to calls to action from within the community: \textit{"There's a lot of petitions for certain movement, or certain things. I was wondering if all of this translates to those thing where you can, or an individual, can do something to drive those changes or those road maps? [Like], they can push their politicians or their local governments to do something towards these goals.''}

\subsection{Tensions between Rigor and Accessibility for User Confidence}

\begin{figure}[ht]
    \centering
    \includegraphics[width=.75\linewidth]{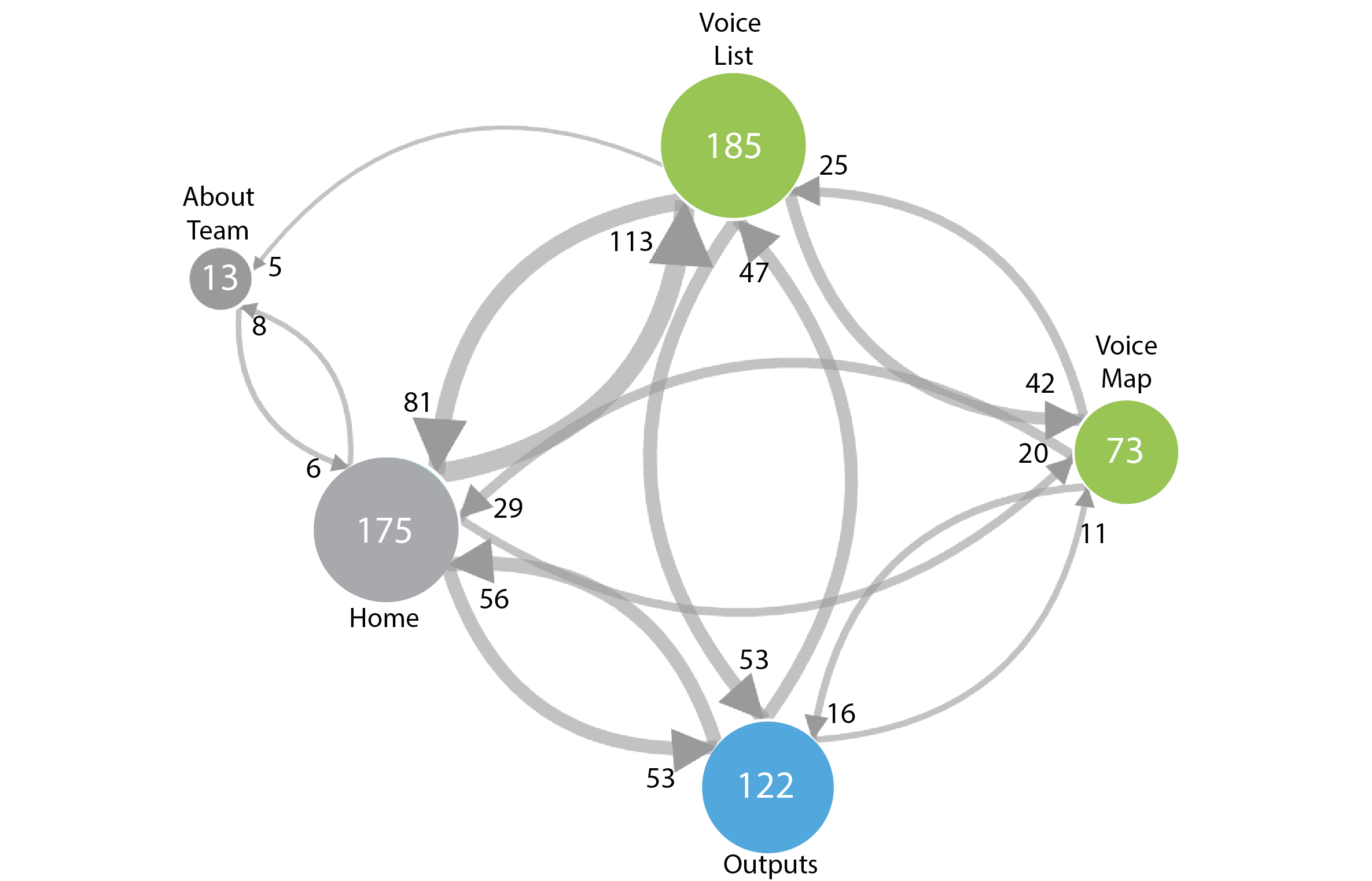}
    \caption{Node-link diagram illustrating how community participants navigated between the five pages in the Voice to Vision community-facing platform: the home page, about team page, voice page with list view, map visualization page, and outputs page. Arrows represent transitions between pages, with the transition counts shown near each arrowhead.}
    \label{fig:node-link}
\end{figure}


While the level of detail and rigor was helpful to promote transparency, the amount of information could be overwhelming for some users~\scaffoldBadge{DG-Scaffold}. For example, C3 explains \textit{''I do think this is a very helpful website in terms of the layout and everything for researchers. But I think if you want to push it to a broader audience, you have to kind of restructure the website.''} C17 wondered if the level of detail takes away from the ability to efficiently and effectively extract information: \textit{``There's so much, and it's so... mind, blowingly cross referenced, which is really awesome.... My big question is like, are there places where the amount of information and choice takes away a little bit from being able to focus on what's important.''} 
While this tool seemed more designed for planners, planner participants were often initially overwhelmed by \emph{``seeing a long list of voices''} (P4).

Well established in the literature is the value of data visualizations to make the complex accessible. Planners and community members found simple bar graphs valuable to make sense of frequency of themes, and how various themes intersected within one goal or strategy. Both similarly emphasized the value of seeing intersection of themes in other ways, like network or cluster graphs. Planners suggested how showing intersections earlier on could support them in identifying \emph{``needs for goals that bring together different areas''} (P2).
while later on, clusters could help as a \emph{``gut-check that developed recommendations align with community feedback''} (P1).


C14 similarly found visualizations to be a more productive, big picture explainer view: \textit{``I think the most accessible information is probably gonna be the most used information. So you wanna give a person a lot of information quickly and easily digestible. If they want to go into the weeds, you guys definitely have the content for them to go into the weeds. And I suspect a good chunk of your users are not gonna.''} C14 and C15 suggested videos, validating the finding that photos and images also supported deeper understanding of the process, and a stronger sense of legitimacy through seeing neighbors on the webpage. 
Further, the map visualization similarly offered a new lens upon the vast data, immediately intriguing community participants who made it there. 

Though participants appreciated the map visualization, not all users navigated to it, as shown in Figure~\ref{fig:node-link}.
There were page 598 transitions logged during the field deployment, and the most common transition was from the home page to the voices page with a list view (113 transitions). 
However, users often got stuck in the list view and rarely made it to visualizations. 

Community participants found the audio of highlights and medleys valuable, improving understanding, legitimacy, and access. Similarly to how photos reflected the community back in an effective and legitimizing way, hearing a voice could do similar work, as C4 notes: \textit{``I like that because you can actually hear how people really feel about the neighborhood instead of just seeing what they type. Because half the time, they can type something fake instead of the actual truth. It's good that we can hear what they're actually saying and all the responses about the neighborhood. And I agree with what they're all saying...''} They continue to deepen their explanation for accessibility: \textit{``Since I'm a person with dyslexia basically somewhere like I can't read or write.''} Further, community members and planners highlighted the importance of translation and mobile-friendly tools early on, and 19.1\% of users interacted with the platform on a mobile or tablet device, though only 4 users (4.8\%) used the Google Translate feature. 
Beyond visualizations and accessibility features, the planners emphasized how the creation of planning goals serves as a way to communicate a neighborhood plan in a more scaffolded way to community members.
As P2 states: \emph{``[goals are] more about connecting the need to the action and is a way to communicate it.''}
Community members resonated with the goals on the outputs page and were curious about them, with 29.8\% of users clicking on at least one goal card to view more details. 
\vspace{-0.28cm}
\subsection{Importance of Adaptable yet Structured Data}

Our iterative design process with planners emphasized the importance of balancing flexibility with consistency across community engagement efforts~\flexiBadge{DG-Flexi}. 
While the underlying data structure was not explicitly presented during user studies, planner participants confirmed its importance when using the sensemaking interface. 
They highlighted the need for flexibility in data management: \emph{``Not everything is informed by outreach - there's also planning rationale and how we approach a study''} (P1). 
Planners valued the system's ability to adapt to specific planning contexts while maintaining structure, noting that \emph{``we need flexibility in how we use community information to develop goals and the package of strategies that support those goals''} (P3).
The sensemaking interface supported this balance by allowing planners to recognize opportunities for additional data collection. 
As P2 explained, \emph{``The system helps us understand where we need to seek more input to develop the package of recommendations,''} demonstrating the balance between flexibility (identifying and integrating new data collection opportunities) and consistency (maintaining a structured framework that makes feedback comparable and traceable across multiple engagement iterations).

Community members also recognized how the system supports the iterative nature of planning work.
Participant C17 described it as \emph{``the repository of all of this knowledge that has been generated, which is now something that I could learn from and use for something else or use to kind of track how the process is going forward, and still, even give additional feedback''}. 
Both stakeholder groups identified opportunities for expanding the system to other contexts, with P5 suggesting it \emph{``could be helpful for applications work... tracking zoning or development patterns''}, 
while C5, a community member, envisioned broader applications: \emph{``it'll be really cool to see this expand for like all the districts.''} 

However, participants highlighted gaps in the current implementation, particularly around demographic data collection and sorting capabilities.
P5 emphasized the need for consistent demographic tracking: \emph{``I want to know more about who provided the feedback... how much they can relate to the neighborhood... Where do they live? What age bracket?''} 
P4 illustrated why this matters: \emph{``What the young people focused was so different from community meetings with adults. Adults talk about parking while young people focus on wanting to get to school on time and finding a place to hang out after school that doesn't cost money.''} 
They further suggested enhanced flexibility in data exploration through additional sort options to \emph{``parse [the] community in different ways.''}

\subsection{Technical Connections for Planners, Trust Connections for Community}

Our system's focus on bridging relationships~\relateBadge{DG-Relate} revealed significant differences in how planners and community members valued different types of connections. 
Planners emphasized the technical relationships between data elements, particularly appreciating visualizations that revealed patterns across community input. 
As P7 noted, the cluster visualization was the \emph{``easiest way for me to process all the information together and understand where the trends are,''} 
while P3 noted that \emph{``having data distributed and clustered on the map is helpful.''} 
Planners consistently described their work as fundamentally about bridging different inputs and perspectives. 
P2 emphasized that \emph{``frequency of mentions is not necessarily an indicator of importance—even if just one person connects dots between issues, planners need the ability to highlight that connection.''} 
The sensemaking interface also revealed unexpected opportunities for collaboration, with P1 noting how topic distributions across strategies helped identify agency partnerships: \emph{``For strategies like fresh food access strategy that goes under health but also has 190 economy comments tagged, it helps clarify interagency collaboration needs.''}

Community members, by contrast, placed less emphasis on navigating technical relationships between data elements. 
Our usage analytics showed limited interaction with features designed to explore these connections, where only 17\% of users clicked between output cards and 19\% of users navigated between voice and output cards, with an average of just 0.3 (+/- 0.8) and 0.5 (+/- 1.3) clicks per user. 
Instead, community members focused more on the human relationships, particularly how the platform built trust between residents and the government. 
As C14 explained, \emph{``they're not gonna personally go through the 3000 [comments], but they wanna know somebody did... that context gives a person the confidence that I might not understand everything on this website, but somebody does and they used it for this reason.''} 
Community members also wanted to see connections between voices and demographic information to assess fairness. 
C9 explained, \emph{``if I see like [ages] from 18 to 80 was taken into consideration, then that makes me know that all ages were taken into consideration, all genders, all neighborhoods... that would make me feel like it was more fair.''} 
These findings suggest that while both stakeholder groups value relationships, they prioritize different types.
Planners focus on analytical connections that inform decision-making, while community members prioritize connections that build trust and demonstrate inclusivity in the planning process.

\section{Discussion}

Our evaluation of the Voice to Vision system reveals its potential in bridging the critical gap between community voices and planning outputs through a structured yet flexible data infrastructure. 
Planners emphasized the system's value in supporting systematic sensemaking, a capability lacking in their typical spreadsheet-based workflows. 
Specifically, the sensemaking interface enabled them to identify connections across diverse inputs and communicate planning outputs in more accessible formats.
Among community members, we observed multifaceted benefits, in which participants valued seeing themselves and others reflected in the voices, discovering patterns within community feedback, and witnessing the rigor behind planning processes. 
However, they consistently emphasized actionability as a critical concern, wanting to see concrete paths from community input to implementation. 
Both stakeholder groups demonstrated a preference for digestible information formats while appreciating the availability of deeper exploration opportunities, a tension we explore further in our discussion of visualization approaches. 
Perhaps most significantly, the system facilitated a sense of legitimacy and care across the planning ecosystem, helping build shared understanding between elected or appointed officials, planners, and community members. 
In the following sections, we reflect on three key themes: the practical insights gained from our participatory design process; the relationship between accessibility, visualization, and user agency in civic platforms; and the potential evolution of Voice to Vision as a \textit{living library} of community voices that preserves institutional memory.

\subsection{Lessons Learned and Recommendations for Participatory Design in Governance}

\subsection{Balancing between Accessibility and User Agency through Visualizations}

\subsection{Voice to Vision as a Living Library of Community Voices}

\section{Conclusion}
Voice to Vision demonstrates how a comprehensive, data-centered sociotechnical system can bridge the persistent gap between community input and planning outcomes, a critical challenge in civic decision-making processes like urban planning. 
Our findings reveal the potential of structured data architectures in fostering shared understanding between stakeholders while supporting both systematic sensemaking for planners and transparency for community members. 
Participants particularly valued seeing themselves reflected in collected voices, witnessing the rigor behind decision-making, and understanding how feedback shaped outcomes. 
However, our work also illuminates persistent tensions around actionability, accessibility, and the balance between digestible overviews and in-depth exploration opportunities. 
Future advances in civic technology may benefit from CSCW research that balances innovative, specialized solutions with investments in reusable and sustainable infrastructure for community engagement.
The most pressing challenges, such as enhancing interoperability between systems, supporting longitudinal engagement across planning cycles, and measuring long-term impact on trust and legitimacy, require commitment to in-situ research partnerships that extend beyond typical project timelines. 
By creating civic platforms designed for sustainability from the outset, we can move beyond transactional engagement to build the enduring sociotechnical infrastructure that meaningful civic participation requires.


\bibliographystyle{ACM-Reference-Format}
\bibliography{main}

\appendix


\section{Planner Design Cohort}

Table~\ref{tab:plannercohort} contains information about each participant in the planner design cohort.

\begin{table*}[ht]
    \scriptsize
    \centering
    \caption{Information about each participant in the planner design cohort
    ``Role'' indicates the participant's professional position. 
    ``Exp (yrs)'' shows the participant's years of experience in urban planning.
    ``Specialty'' describes the participant's area of planning expertise. 
    ``Session Format'' indicates whether the participant took part in the workshop or individual interview session. 
    Participants P1 to P3 participated in the planner user studies.
    Participants P2 (*) and P16 (*) are co-authors of this paper.}
    \begin{tabular}{@{}p{0.07\linewidth}|p{0.35\linewidth}|p{0.1\linewidth}|p{0.35\linewidth}}
        \toprule
        \textbf{ID}&\textbf{Role}&\textbf{Exp (yrs)}&\textbf{Specialty}\\
        \midrule
        P1&Senior Lead&12&Urban Design\\
        P2 (*)&Director&20&Civic Engagement\\
        P3&Senior Planner&7&Housing\\
        P8&Chief Operating Officer&20&Agency Operations, Urban Planning\\
        P9&Planner&5&Housing\\
        P10&Deputy Director&8.5&Neighborhood Planning\\
        P11&VP Government \& Community Relations&15&Economic Development\\
        P12&Director&21&Policy Development\\
        P13&Assistant Director&15&Capital Planning\\
        P14&Deputy Director&10&Neighborhood Planning\\
        P15&Staff Designer&3&Civic Engagement\\
        P16 (*)&Senior Lead&6&Civic Engagement, Neighborhood Planning\\
        \bottomrule
    \end{tabular}
    \label{tab:plannercohort}
\end{table*}

\section{Community Design Cohort}

Table~\ref{tab:communitycohort} contains information about each participant in the community design cohort.

\begin{table*}[ht]
    \scriptsize
    \centering
    \caption{Information about each participant in the community design cohort.
    ``ID'' refers to the participant identifier.
    ``Experience'' summarizes the participant's experience with past planning efforts.
    ``Background'' describes additional information included in the participant's responses to the open call.
    ``Role'' indicates whether the participant was in the Design Team or Advising Team.
    Participants C5, C6, and C10 participated in the community user studies.}
    
    \begin{tabular}{@{}p{0.06\linewidth}|p{0.3\linewidth}|p{0.4\linewidth}|p{0.1\linewidth}}
        \toprule
        \textbf{ID}&\textbf{Experience}&\textbf{Background}&\textbf{Role}\\
        \midrule
        C5&Attended Community Board meetings&Youth (16-24 years old) and works in a civic organization&Design\\
        C6&Attended Community Board meetings&Bi-lingual speaker&Design\\
        C10&Attended Community Board meetings&Long time resident and community leader&Advising\\
        C18&No experience&Youth (16-24 years old)&Design\\
        C19&No experience&Bi-lingual speaker and works in a non-profit&Advising\\
        C20&Attended Community Board meetings&Bi-lingual speaker and community leader&Advising\\
        C21&Participated in previous planning project&Artist&Advising\\
        C22&No experience&Software Engineer&Design\\
        C23&Attended Community Board meetings&Youth (16-24 years old) and works in a civic organization&Design\\
        \bottomrule
    \end{tabular}
    \label{tab:communitycohort}
\end{table*}

\end{document}